\documentclass[10pt,aps,amsfonts,amssymb,floatfix,superscriptaddress,notitlepage,footinbib]{revtex4-1}
\usepackage{mathrsfs}
\usepackage{bm}
\usepackage{bbm}
\usepackage{multirow}
\usepackage{graphicx,color}

\usepackage[normalem]{ulem}
\usepackage{lineno}
\usepackage{tabularx}
\usepackage{physics}                                                            
\usepackage[colorlinks=true,linkcolor=blue,filecolor=magenta,urlcolor=blue,citecolor=blue]{hyperref}

\newcommand{\nnum}{\nonumber}
\def\angle<#1>{\langle {#1} \rangle}
\def\qangle<#1>{\left\langle {#1} \right\rangle}
\def\sub.#1|{\left.{#1}\right|}

\newcommand{\blue}[1]{\textcolor{black}{#1}}

\renewcommand{\Vec}[1]{{\bf #1}}

\newcommand{\beq}{\begin{equation}} \newcommand{\eeq}{\end{equation}}
\newcommand{\beqn}{\begin{eqnarray}} \newcommand{\eeqn}{\end{eqnarray}}

\newcommand{\eq}[1]{Eq.~(\ref{#1})}

\begin{document}

\title{
Replicated liquid theory
in \texorpdfstring{$1+\infty$}{TEXT} dimensions
}

\author{Yukihiro Tomita}
\affiliation{Department of Physics,  Osaka University, Toyonaka, Osaka 560-0043, Japan}

\author{Hajime Yoshino}
\affiliation{D3 Center, Osaka University, Toyonaka, Osaka 560-0043, Japan}
\affiliation{Department of Physics,  Osaka University, Toyonaka, Osaka 560-0043, Japan\\
}
\email{yoshino.hajime.cmc@osaka-u.ac.jp}

\begin{abstract}
We develop a replicated liquid theory for structural glasses 
that exhibit spatial variation
of physical quantities along one axis, say $z$-axis.
The theory becomes exact with infinite transverse dimension $d-1 \to \infty$.
It provides an exact free-energy functional with space-dependent glass
order parameter $\Delta_{ab}(z)$.
As a first application of the scheme, we study 
diverging lengths associated with dynamic/static glass transitions of hard-spheres
with/without a confining cavity.
The exponents agree with those obtained in previous studies on related mean-field models. Moreover, it predicts a non-trivial spatial profile of the glass order parameter $\Delta_{ab}(z)$ within the cavity, which exhibits a scaling feature approaching the dynamical glass transition.
\end{abstract}
\maketitle

\tableofcontents

\section{Introduction}

Recently the exact mean-field theory for super-cooled, glass forming liquids
which becomes exact in the large dimensional limit $d\to \infty$ was established
\cite{kurchan2012exact,kurchan2013exact,charbonneau2014exact,charbonneau2014fractal,
  YZ14,rainone2015following,1130848328245474560,urbani2023gardner}
\blue{
  which is based on the density functional theory of liquids \cite{hansen1990theory}
  and the replica method developed in the studies of disordered systems \cite{MPV87}.}
This is a significant theoretical achievement since the key notions of glass physics
\cite{kirkpatrick1987stable,kirkpatrick1987connections,angell2000relaxation,cavagna2009supercooled,berthier2011theoretical}
conceived by preceding experimental, numerical and theoretical studies
such as the dynamical glass transition, thermodynamic glass transition (Kauzmann transition),
jamming and yielding were firmly established and understood in a unified manner in a single theoretical framework based on 1st principles.
The central object is the glass order parameter
$\Delta_{ab}$ which parameterize the relative mean-squared displacements among replicated
liquids $a=1,2,\ldots$. It remains to be clarified to what extent the features
established in the large dimensional limit, such as the very existence of the Kauzmann transition, 
remain valid in finite dimensional real systems.

An obvious drawback of $d\to \infty$ theories is that, by its construction, it cannot describe any spatial variation or fluctuation of physical quantities. 
Actually glasses forming liquids and glasses are known to exhibit various interesting spatial heterogeneities such
as the dynamical heterogeneity observed in the supercooled liquid state
\cite{karmakar2014growing,biroli2008thermodynamic,bouchaud2004adam},
the isostatic length which diverges approaching jamming \cite{wyart2005rigidity}
and formation of shear-bands approaching yielding or fracture \cite{maloney2006amorphous}.
Aiming to capture these spatial heterogeneities theoretically,
we develop an exact mean-field theory for $1+\infty$ dimensional system
so that we become able to describe the spatial variation and fluctuation of physical properties
along one spatial axis, say $z$-axis. The main object is the space dependent glass order parameter
$\Delta_{ab}(z)$. Our approach is related but somewhat different from the usual Ginzburg-Landau (GL) type, field theoretic descriptions \cite{lubensky2023renormalization}. While such GL approaches will become useful at sufficiently long-wave lengths, our theory is derived microscopically so that it is precise also at the particle scales. We believe it will become particularly useful in situations like jamming
and yielding where accurate microscopic descriptions at the scale of particles are indispensable.

The purpose of this paper is twofold. First we develop a generic replicated liquid theory
in $1+(d-1)$ dimensions which becomes exact in $d-1 \to \infty$ limit.
Second we test the scheme by analyzing the length scales which diverge
approaching the dynamic/static glass transitions using hard-spheres as the simplest glass forming system.

The analysis of the diverging length scales is done in two setups.
First setup is an in an infinitely large system $-\infty < z < \infty$.
There we analyze the spatial correlation of the thermal fluctuations of the glass order parameter. 
We find a length scale which diverges approaching the dynamical glass transition.
It was originally predicted \blue{by a study on a mean-field spin-glass model \cite{kirkpatrick1987stable}
whose phenomenology closely resemble that of structural glasses \cite{kirkpatrick1987connections}.}
Later the inhomogeneous MCT (mode coupling theory) \cite{biroli2006inhomogeneous} obtained the
length scale from a 1st principle.
The 2nd setup is a cavity system of finite depth $L$ containing the hard-spheres. 
This setup allows one to study the correlation 
lengths, called as the point-to-set lengths in the literature
\cite{biroli2008thermodynamic,bouchaud2004adam},
which diverge approaching the two glass transitions: the dynamical transition already mentioned above
and the static glass transition or the Kauzmann transition.
We capture the correlation lengths through the following two features.
One is the finite size effect on the dynamic/static length glass transitions:
we study how the glass transition points are affected by the finiteness of the cavity size $L$. 
The other is the spatial variation of the glass order parameter $\Delta_{ab}(z)$
viewed as a function of the distance from the cavity wall.
It turned out to be very similar to 
the behavior of the order parameter associated with the surface critical phenomena
\cite{binder1972phase,kikuchi1985monte1}.
Our results reconfirm the critical exponents obtained in previous theoretical studies based on
mean-field spin-glass model \cite{kirkpatrick1987stable}, 
inhomogeneous MCT \cite{biroli2006inhomogeneous}, a Kac glass model \cite{franz2007analytic}
and a one-dimensional chain of discrete cells containing a hard-sphere liquid \cite{ikeda2015one}.
Our analysis can be viewed  as a thermodynamic (static) counterpart of
the inhomogeneous MCT \cite{biroli2006inhomogeneous}
and a continuous limit of the chain model \cite{ikeda2015one}.
Our work is also motivated in part by 
the replica theory of supervised learning by a prototypical deep neural network
which revealed a non-trivial spatial profile of the glass order parameter \cite{yoshino2020complex}.

\section{Model}

\begin{figure}[t]
\begin{center}
  \includegraphics[width=8cm]{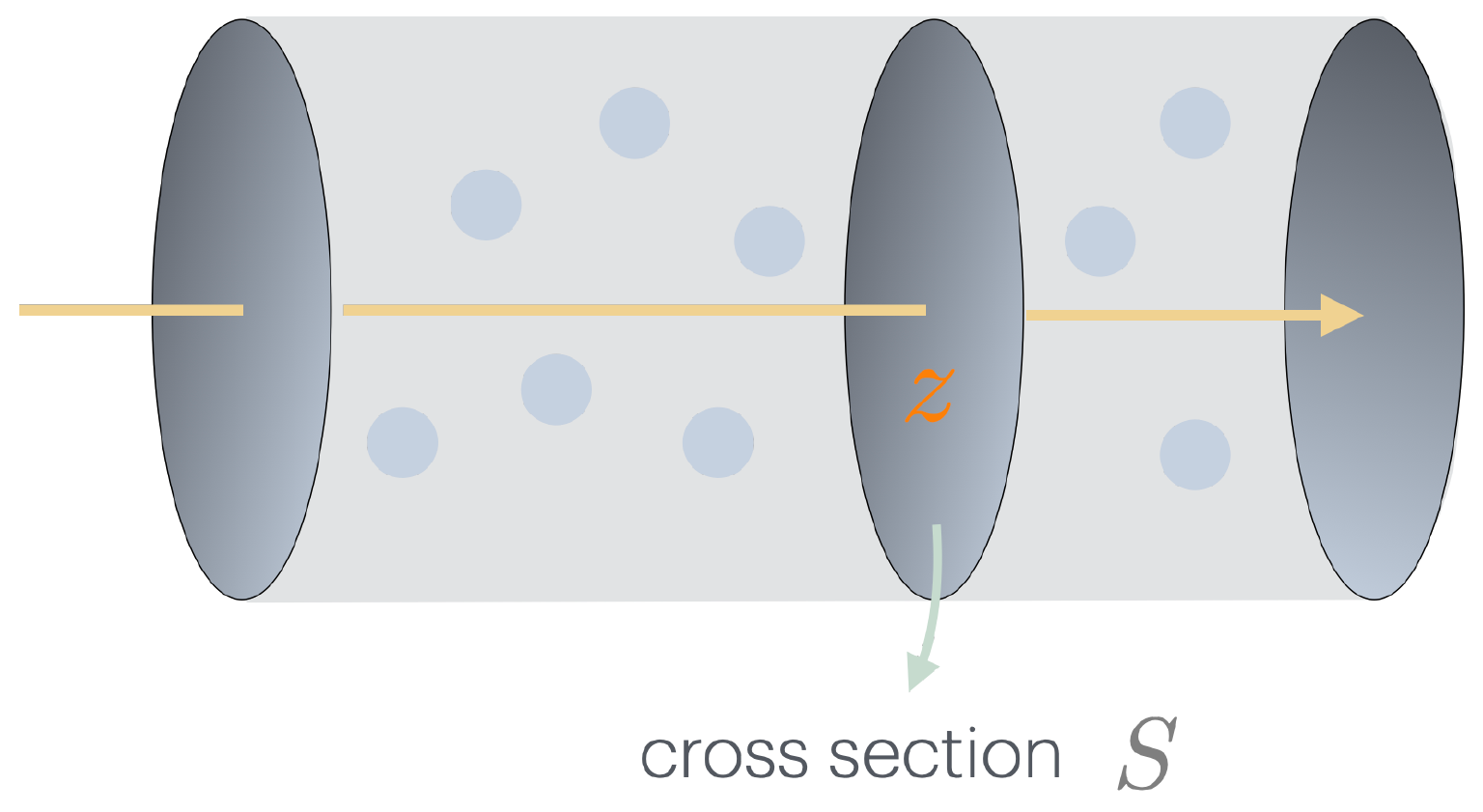}
\end{center}
\caption{
Schematic picture of our system in a cylinder
}
\label{fig_cylinder}
\end{figure}

We consider an assembly of particles $i=1,2,\ldots,N$ of mass $m$ contained in a cylinder of
cross-section area $S$ as shown in  Fig.~\ref{fig_cylinder}.
The coordinates of the particles are given by $\Vec{x}_{i}=(x_{1 i},x_{2 i},\ldots,x_{d-1 i},z_{i})$
while their momentum are $\Vec{p}_{i}$.
The particles are interacting with each other through a two-body interaction potential $v(r_{ij})$
with $r_{ij}=\qty|\Vec{x}_{i}-\Vec{x}_{j}|$. Then the Hamiltonian is given by.
\begin{align}
    \label{hamiltonian_of_liquid_theory}
    H=\sum_{i=1}^{N}\frac{\qty|\bm{p}_{i}|^{2}}{2m}+\sum_{i<j}v(r_{ij})+\sum_{i=1}^{N} U(z_{i})
\end{align}
The last term represents a potential which can be used, for example, to confine the particles
within a finite region along the z-axis. We are interested in static macroscopic properties of the system
in equilibrium at temperature $T$ or inverse temperature $\beta=1/k_{\rm B}T$ with $k_{\rm B}$ being the Boltzmann constant.

As a specific model system we will consider hard-spheres (HS) with diameter $D$
with the interaction potential given by,
\beq
v(r)=\left \{ \begin{array}{ll}
\infty & (r\leq D) \\
0 & (r>D)
\end{array} \right.
\eeq
Thus the Boltzmann factor associated with the HS potential  becomes $\exp(-\beta v(r))=\theta(r)$
\blue{where $\theta(r)$ is the Heaviside step function.}

\section{Construction of an inhomogeneous replicated liquid theory}

In this section, we discuss the construction of an inhomogeneous replicated liquid theory.
To this end  we first construct a density functional theory for simple liquids in $1+(d-1)$ dim space which becomes exact in $d-1 \to \infty$ limit. We obtain an exact form of the
free-energy functional expressed in terms
of density profile $\rho(z)$ which is allowed to vary along the $z$-axis.
Next, we replicate the system and construct an inhomogeneous replicated liquid theory with the free-energy
expressed exactly in terms of the density profile $\rho(z)$ and space dependent glass order parameter $\Delta_{ab}(z)$.
Then we derive the self-consistency equation which determines
the density profile and the glass order parameter.
We show how various thermodynamic quantities including chemical potential, pressure, and
structural entropy (complexity) can be computed.
The details of the derivations are shown in appendix
\ref{appendix-liquid-theory},
\ref{appendix-replicated-liquid-theory} and
\ref{appendix-1RSB}.

\subsection{Inhomogeneous liquid theory}

We assume that the system is uniform within each cross-section of the cylinder 
and the density varies only along the $z$ axis \blue{(see Fig.~\ref{fig_cylinder})}.
Then we naturally
introduce a microscopic density profile,
\beq
\rho(z)=
\sum_{i=1}^{N} \langle \delta(z-z_{i}) \rangle
\eeq
where $\langle \ldots \rangle$ is the thermal average.
\blue{Although the density functional theory for liquids is well established
  \cite{hansen1990theory}, we show in appendix \ref{appendix-liquid-theory-basic-setup} the derivation of the density functional expressions
for the sake of clarity and pedagogical reasons.}
The free-energy of the system as a functional of the density profile $\rho(z)$ is obtained as,
\beqn
\frac{-\beta F[\rho]}{S} &=&
\int_{-\infty}^{\infty} dz \rho(z) [ 1-\log (\lambda_{\rm th}^{d} \rho(z))] + \int dz \rho(z) (-\beta U)(z) \nonumber \\
&& +\frac{1}{2}\int_{-\infty}^{\infty} dz_{1} \int_{-\infty}^{\infty} dz_{2}
\rho(z_{1})\rho(z_{2})
\frac{1}{S}\int \prod_{\mu=1}^{d-1} d x_{\mu 1}
\int \prod_{\mu=1}^{d-1} d x_{\mu 2} f(r^{2}_{12})
\eeqn
where $\lambda_{\rm th}=h/\sqrt{2\pi m k_{\rm B}T}$ is the thermal de Broglie wave length and
\beq
f(r^{2})=e^{-\beta v(r)}-1
\eeq
is the Mayer function and
\beq
r^{2}_{12}=(z_{1}-z_{2})^{2}+\sum_{\mu=1}^{d-1}(x_{\mu 1}-x_{\mu 2})^{2}.
\eeq
In the large dimensional limit $d-1 \to \infty$ contributions from higher orders in the
Mayer expansion becomes negligible \cite{FP99,1130848328245474560} and the above expression becomes exact.

The equilibrium density profile $\rho(z)$ 
is the one which minimizes the free-energy functional $F[\rho]$
under the constraint,
\beq
\frac{N}{S}=\int_{-\infty}^{\infty} dz \rho(z)
\label{eq-sum-rule-density}
\eeq
It can be obtained by solving
\beqn
0&=&\frac{\delta}{\delta \rho(z)}
\left ( \frac{-\beta F[\rho]}{S}+\beta \mu \int dz \rho(z)\right).  
\eeqn
Here we introduced a Lagrange parameter $\beta \mu$ where
$\mu$ can be interpreted as the chemical potential.
The above equation yields $\mu$ which must be adjusted to satisfy \eq{eq-sum-rule-density}.

\blue{In appendix \ref{appendix-liquid-theory-large-dim-limit} we explain
  a rescaled length scale (see \eq{eq-def-hat-z})
  and a rescaled density (see \eq{eq-def-hat-var-phi})
  which we will use below. There we also present
  the free-energy functional and the chemical potential
  expressed in terms of these rescaled quantities.
  In appendix \ref{sec-compression-liquid} we consider compression of the
  system to obtain an expression of pressure.
  }

\subsection{Inhomogeneous replicated liquid theory}

\blue{Next let us consider inhomogeneous replicated liquid theory aiming to extend
  the conventional theory for the homogeneous system \cite{1130848328245474560}.
We considers system of $m$-replicas $a=1,2,\ldots,m$ all obey
the same Hamiltonian \eq{hamiltonian_of_liquid_theory}.
The basic idea of the cloned or replicated liquid theory  \cite{mezard1999first} is to consider the possibility
of spontaneous formation of a 'molecular liquid state' with molecules made of 'replicas'.
To describe such a molecular liquid state, it is convenient to decompose
the coordinate  $\bm{x}_{i}^{a}$ of particle $i$ in the $a$ replica as,}
\begin{align}
    \bm{x}_{i}^{a}
    =(\bm{x}_{i})_{\rm c}+\bm{u}_{i}^{a}
    \label{eq-decomposition-coordinate-main}
\end{align}
with centers of the 'molecules' $i=1,2,\ldots,N$ located at
\beq
(\bm{x}_{i})_{\rm c}=((x_{i}^{1})_{\rm c},(x_{i}^{2})_{\rm c},\ldots, (x_{i}^{d-1})_{\rm c},(z_{i})_{\rm c})=\frac{1}{m}\sum_{a=1}^{m}\bm{x}_{i}^{a} 
\eeq
The 2nd term in the r.h.s of \eq{eq-decomposition-coordinate-main}
represents thermal fluctuations within the molecules.
We introduce the space dependent glass order parameters as,
\beq
\Delta_{ab}(z)=\alpha_{aa}(z)+\alpha_{bb}(z)-2\alpha_{ab}(z) \qquad 
\alpha_{ab}(z) \rho(z) =
\frac{d}{D^{2}}
\frac{1}{S}
\sum_{i=1}^{N} \langle {\bm u}_{i}^{a} \cdot {\bm u}_{i}^{b} \delta(z-(z_{i})_{\rm c})
\rangle
\label{eq-def-order-parameters}
\eeq
where $a$,$b$ are indices for replicas $a,b=1,2,\ldots,m$.
\blue{Formation of a molecular liquid state is signaled by the presence of finite glass order parameter
$0  < \Delta_{ab} < \infty$.   Physically this means emergence of a glassy state
  while $\Delta_{ab}=\infty$ in a purely liquid state
    and $\Delta_{ab}=0$ in a jammed state.
  }

It has been realized that the relevant length scale for the fluctuation of the inter-particle distance
is $O(1/\sqrt{d})$ in the large dimensional limit $d\to \infty$ \cite{kurchan2012exact,1130848328245474560}
\blue{(see appendix \ref{appendix-liquid-theory-large-dim-limit})}.
This naturally leads us to introduce a scaled coordinate $\hat{z}$ as,
\beq
z=\frac{D}{\sqrt{d}}\hat{z}
\label{eq-def-hat-z}
\eeq
Here $D$ is the microscopic length scale which characterize the interactions, which is the diameter of spheres in the case of hard-spheres.

As we explain in the appendix \ref{sec-basic-setup-replicated-liquid-theory},
the free energy functional of the replicated system is obtained exactly in the large dimensional limit $d-1 \to \infty$
\blue{using the glass order parameter introduced above} as,
\begin{align}
    &\frac{-\beta F_{m}[\rho, \alpha_{ab}]}{S}\frac{\sqrt{d}}{D}\nnum\\
  &=\int^{\infty}_{-\infty} d\hat{z} \rho(\hat{z})\qty{1-\ln{(\rho(\hat{z}) \lambda_{\rm{th}}^{d})} + d\ln{m}+\frac{(m-1)d}{2}\ln{\frac{2\pi e (D/\lambda_{\rm{th}})^2}{d^2}+\frac{d}{2}\ln{\det\qty((\alpha(\hat{z}))^{m,m})}}+ (-\beta U)(\hat{z}) }
  \nnum\\
    &+\frac{d}{2}\frac{\Omega_{d}D^d}{d\sqrt{2\pi}}
    \int^{\infty}_{-\infty} d\hat{z} \rho(\hat{z})
    \int^{\infty}_{-\infty} d\hat{z}' \rho(\hat{z}')
    e^{-\frac{(\hat{z}-\hat{z}')^2}{2}}
    \qty(-\mathcal{F}_{\rm{int}}(\Delta_{ab}(\hat{z},\hat{z}'))).
\end{align}
Here, $-\mathcal{F}_{\rm{int}}$
\blue{which represents the replicated Mayer function},
is defined as
\begin{align}
    -\mathcal{F}_{\rm{int}}(\Delta_{ab}(\hat{z},\hat{z}'))
    =\int^{\infty}_{-\infty}d\xi e^{\xi}
    e^{-\frac{1}{2}\sum_{ab}\Delta_{ab}(\hat{z},\hat{z}')\partial_{\xi_a}\partial_{\xi_b}}\left.
    \qty[\prod_{a}e^{-\beta v\qty(D^2\qty(1+\frac{\xi_a}{d})^2)}-1]
    \right|_{\qty{\xi_a=\xi}}.
\end{align}
\blue{We also introduced a short-hand notation},
\beq
\Delta_{ab}(\hat{z}_{1},\hat{z}_{2})=\frac{\Delta_{a}(\hat{z}_{1})+\Delta_{b}(\hat{z}_{2})}{2}
\eeq

The equilibrium density profile $\rho(\hat{z})$ is the one which minimizes the free-energy under the constraint
\eq{eq-sum-rule-density-hat} (see below).
The density profile $\rho(\hat{z})$ and glass order parameters $\Delta_{ab}(\hat{z})$ are obtained by solving
\beqn
0&=&\frac{\delta}{\delta \rho(\hat{z})}
\left [ \frac{(-\beta F_{m}[\rho,\alpha_{ab}])}{SD/\sqrt{d}} +\beta \mu 
  \int d\hat{z} \rho(\hat{z}) \right] \nonumber \\
0&=& \frac{\delta}{\delta \alpha_{ab}(\hat{z})}
\frac{(-\beta F_{m}[\rho,\alpha_{ab}])}{SD/\sqrt{d}}
\eeqn
The 1st equation yields the chemical potential $\mu$ which must be adjusted to satisfy \eq{eq-sum-rule-density}
which reads as
\beq
\frac{N}{S}=\frac{D}{\sqrt{d}}\int_{-\infty}^{\infty} d\hat{z} \rho(\hat{z})
\label{eq-sum-rule-density-hat}
\eeq

In the case of hard-sphere like systems, $D$ is the radius of particles. For those cases
it is convenient to introduce the volume fraction $\varphi(\hat{z})$,
\beq
    \varphi(\hat{z})
    =\rho(\hat{z})\Omega_{d}\qty(\frac{D}{2})^{d}
    \eeq
    where $\Omega_{d}$ is the volume of $d$-dimensional unit sphere,
In order to study glassy states in the large dimensional limit it becomes convenient
to introduce a scaled volume fraction \cite{1130848328245474560}
    \blue{(see appendix \ref{appendix-liquid-theory-large-dim-limit})},
\beq
    \hat{\varphi}(\hat{z})
    \equiv\frac{2^{d}\varphi(\hat{z})}{d}
    =\frac{\rho(\hat{z})\Omega_{d} D^{d}}{d}.
    \label{eq-def-hat-var-phi}
    \eeq

    \blue{In appendix \ref{appendix-liquid-theory-large-dim-limit}
      we also report an expression of the chemical potential,
      and the free-energy functional for the bulk (uniform system) for references.
 In appendix \ref{sec-compression-replica} we consider
      compression of the system and obtain an expression of the pressue.}

\subsection{One Step RSB solution}

The simplest ansatz for the matrix form of the glass order parameter
\blue{defined in \eq{eq-def-order-parameters}}
is
\beq
\alpha_{ab}(\hat{z})=(m \delta_{ab}-1)\alpha(\hat{z})
\eeq
or
\beq
\Delta(\hat{z})=2m\alpha(\hat{z}) \qquad \Delta_{ab}(\hat{z})=\Delta(\hat{z})(1-\delta_{ab})
\eeq
which are symmetric under permutations of replicas $a=1,2,\ldots,m$.
This symmetry is the so-called replica symmetry.
However we call these ansatz as one step RSB (1RSB) ansatz in the present paper
for the reasons we explain in appendix \ref{appendix-1RSB} \blue{and \ref{sec-basic-setup-replicated-liquid-theory}}.
\blue{This ansatz is known to be valid at densities lower than the so called Gardner transition density \cite{kurchan2013exact,urbani2023gardner}
  in the uniform system.}

Using this ansatz we can evaluate thermodynamic quantities as we explain in appendix \ref{appendix-1RSB}.
The free-energy is obtained as
\beqn
    &\frac{-\beta F_{m}^{\rm{1RSB}} \qty[\qty{\Delta(\hat{z})}]}{S}\frac{\sqrt{d}}{D}\frac{\Omega_{d}D^{d}}{d}
    =
    \int d\hat{z} \hat\varphi(\hat{z})
    \qty{1-\ln{(\hat\varphi(\hat{z}) (d/\Omega_{d})(\lambda_{\rm{th}}/D)^{d})}
      + d\ln{m}+\frac{(m-1)d}{2}\ln{\frac{2\pi e D^2}{d^2\lambda_{\rm{th}}^{2}}}
      +(-\beta m \hat{U}_{0}(\hat{z}))
    },
\nnum\\
    +&\frac{d}{2}
    \qty{
     \int d\hat{z} \hat\varphi({z})
     \qty[2(-\beta m \hat{U}_{1}(\hat{z}))
       +(m-1)\ln{\frac{\Delta(\hat{z})}{2}}-\ln{m}]
    +
    \int d\hat{z}_{1} \hat\varphi(\hat{z}_{1})
    \int d\hat{z}_{2} \hat\varphi(\hat{z}_{2})
    \frac{e^{-\frac{(\hat{z}_{1}-\hat{z}_{2})^{2}}{2}}}{\sqrt{2\pi}}
    \qty(-\mathcal{F}_{\rm{int}}\qty(\Delta(\hat{z}_{1},\hat{z}_{2})))}
    \eeqn
    with \blue{the replicated Mayer function given by},
    \beq
    -\mathcal{F}_{\rm{int}}(\Delta)=\int^{\infty}_{-\infty}d\xi
    e^{\xi-\frac{1}{2}\Delta(\hat{z},\hat{z}')}
    \qty[g^m\qty(\xi,\Delta(\hat{z},\hat{z}'))-1]
    \label{eq-F-int-1RSB}
    \eeq
where
    \beq
g\qty(\xi,\Delta) = e^{\frac{1}{2}\Delta\partial_{\xi}^{2}}e^{-\beta v(D^2(1+\xi/d)^2)}
=\int \mathcal{D}w e^{-\beta v\qty(D^2\qty(1+\frac{\xi+\sqrt{\Delta}w}{d})^2)}
\eeq
\blue{Here we used the identity
  \eq{eq-diff-gaussian-integral-conversion}
and  a short-hand notation
  \eq{eq-def-Dz},  $\int \mathcal{D} z \ldots = \int^{\infty}_{-\infty}\frac{dz}{\sqrt{2\pi}} e^{-\frac{z^{2}}{2}}\ldots$.
  }

By  taking a functional derivative of the free energy by $\Delta(\hat{z})$,
we obtain the self-consistent equation for $\Delta(\hat{z})$,.
\begin{align}
    \frac{1}{\Delta(\hat{z})}
    &=\frac{m}{2}\int^{\hat{L}}_{0} \frac{d\hat{z}'}{\sqrt{2\pi}}
    \hat{\varphi}(\hat{z}')
    e^{-\frac{(\hat{z}-\hat{z}')^2}{2}}
    \int_{-\infty}^{\infty} d\xi e^{\xi-\frac{1}{2}\Delta(\hat{z},\hat{z}')}g^{m}(\xi,\Delta(\hat{z},\hat{z}'))\qty(f'(\xi,\Delta(\hat{z},\hat{z}')))^{2}.
\label{self_consistent_eq_most_general}
\end{align}
where
\beq
f(\xi,\Delta)=-\ln{g(\xi,\Delta)}
\eeq

\section{Diverging length scales at glass transitions}

Now let us test our theoretical framework analyzing diverging length scales at dynamic/static glass transitions
of hard-spheres in two different setups.

\subsection{Spatial correlation of glassy fluctuations around the dynamical transition}
\label{sec-spatial-correlation-of-glassy-fluctuation-around-dynamical-transition}

In the first setup we consider an infinitely large system $-\infty < z < \infty$ and examine the
spatial correlation function of the fluctuation of the glass order parameter
around the equilibrium value $\Delta$,
\beq
\delta \Delta(\hat{z})=\Delta(\hat{z})-\Delta.
\label{eq-def-fluctuation}
\eeq

This is done by analyzing the Hessian matrix.
As explained in appendix \ref{sec-longtidudinal-mode-of-hessian} we obtain the Hessian matrix as
\beqn
-M(\hat{z}_{1},\hat{z}_{2})&=&\frac{\partial}{\partial \Delta(\hat{z}_{1})}\frac{\partial}{\partial \Delta(\hat{z}_{2})}
\frac{-\beta F_{m}^{\rm{1RSB}} \qty[\qty{\Delta(\hat{z})}]}{S D/\sqrt{d}} \frac{\Omega_d D^{d}}{d}\nonumber \\
&=& \frac{d}{2}(m-1)\hat{\varphi}(\hat{z}_{1})
\left[
  -\frac{1}{\Delta(\hat{z}_{1})^{2}}\delta(\hat{z}_{1}-\hat{z}_{2})
  +\frac{m}{2}\int d\hat{z} \hat{\varphi}(\hat{z})\frac{e^{-\frac{(\hat{z}_{1}-\hat{z})^{2}}{2}}}{\sqrt{2\pi}}
  \frac{\delta(\hat{z}_{1}-\hat{z}_{2})+\delta(\hat{z}-\hat{z}_{2})}{2}X(\Delta(\hat{z}_{1},\hat{z}))
  \right]\qquad
\eeqn
with
\beqn
X(\Delta)&=&-\frac{\partial}{\partial \Delta}\int d\xi e^{-\frac{\Delta}{2}\frac{\partial^{2}}{\partial \xi^{2}}}
g^{m}(\xi,\Delta)(f'(\xi,\Delta))^{2} \nonumber \\
&=& \frac{1}{2} \int d\xi e^{-\frac{\Delta}{2}\frac{\partial^{2}}{\partial \xi^{2}}}
\left[2(f^{''}(\xi))^{2}+(m-1)(-4f''(\xi))(f'(\xi))^{2}+m(m-1)(f'(\xi))^{4}\right]
\eeqn

By performing Fourier transform we find
\beq
\hat{M}(k)=\int \frac{d\hat{z}}{\sqrt{2\pi}}e^{ik\hat{z}}M(\hat{z})
=\frac{d}{2}(m-1)\hat{\varphi}\left[M_{0}+\frac{k^{2}}{2}M_{2}+O(k^{4})\right]
\eeq
with
\beq
M_{0}=\frac{1}{\Delta^{2}}-\frac{m}{2}\hat{\varphi}X(\Delta) \qquad
M_{2}=\frac{m}{4}\hat{\varphi}X(\Delta) 
\eeq
From the above result we immediately find \blue{the spatial correlation function
the local mean-squared displacements \eq{eq-def-fluctuation} as}, 
\beq
\langle \delta \Delta(\hat{z}_{1})\delta \Delta(\hat{z}_{2}) \rangle
\propto \exp \left(-\frac{|\hat{z}_{1}-\hat{z}_{2}|}{\xi^{\rm hessian}_{\rm d}}\right)
\label{eq-correlation-function-of-Delta}
\eeq
with the correlation length $\xi$ given by
\beq
\xi^{\rm hessian}=\sqrt{\frac{M_{2}}{2M_{0}}} 
\label{eq-correlation-length-from-Hessian-0}
\eeq

\blue{ Now let us focus on the dynamical transition at $\hat\varphi_{\rm d}$.
As we recall in appendix \ref{sec-dynamical-transition-in-uniform-system},
the dynamical transition takes place at 
the spinodal point where the saddle point solution disappears \cite{kirkpatrick1987stable}.
Approaching $\hat\varphi_{\rm d}$ we find the correlation length behaves,}
\beq
\xi^{\rm hessian}_{\rm d}=\sqrt{\frac{M_{2}}{2M_{0}}}  \propto
\delta\hat\varphi^{-1/4}
\label{eq-correlation-length-from-Hessian}
\eeq
with
\beq
\delta\hat\varphi=(\hat\varphi-\hat\varphi_{\rm d})/\hat\varphi_{\rm d}
\label{eq-def-delta-hat-varphi}
\eeq
which measures the distance to the critical point $\hat\varphi_{\rm d}$.
Here we used the fact that $M_{0}$ is nothing but the Hessian of the bulk system
which scales as
\beq
M_{0} \propto \delta\hat\varphi^{1/2}
\eeq
close to the dynamical transition density $\hat\varphi_{\rm d}(m)$
\blue{(See \eq{eq-scaling-M0})} 
\blue{This scaling reflects the fact that the dynamical transition takes place at a spinodal point \cite{binder1987theory,kirkpatrick1987stable}.}
On the other hand $M_{2}$ is essentially a constant close to the critical point.

\subsection{Glass transitions within cavities}

Now we turn to our 2nd setup which is a cavity system.
\blue{The techinical details are reported in appendix \ref{appendix-cavity}.}
It is prepared as follows.
We consider again an infinitely large system $-\infty < \hat{z} < \infty$ with uniform
density $\hat\varphi(\hat{z})=\hat\varphi$. Suppose that the entire system is in the liquid
state. Then we freeze-out the system setting $\Delta(\hat{z})=0$ everywhere except for the 'cavity' region
$0 < \hat{z} < \hat{L}_{\rm cav}$. The free energy of such a cavity system
within the 1RSB anasatz is given by
\beqn
&& \frac{-\beta F_{m}[\hat\varphi,q_{ab}]}{SD/\sqrt{d}}\frac{\Omega_{d}D^{d}}{d}=
\hat{L}_{\rm{cav}} \hat{\varphi}
\left [1-
  \ln \left (\hat{\varphi}(d/\Omega_{d})(\lambda_{\rm{th}}/D)^{d} \right)
 + d\ln{m}+\frac{(m-1)d}{2}\ln{\frac{2\pi e D^2}{d^2\lambda_{\rm{th}}^{2}}}
  \right]
\nonumber
\\
&& +\frac{d}{2} \hat{\varphi}\left\{  \int_{0}^{\hat{L}_{\rm cav}} d\hat{z}
\hat\varphi(\hat{z}) \left[ (m-1)\ln{\frac{\Delta(\hat{z})}{2}}-\ln{m}\right]
       \right.
  \nonumber \\
&& \left.    +\hat\varphi
    \qty[\int^{\infty}_{-\infty} d\hat{z}_{1}\int^{\infty}_{-\infty} d\hat{z}_{2}-\int_{\rm{ex-cav}} d\hat{z}_{1}\int_{\rm{ex-cav}} d\hat{z}_{2}]
    \frac{e^{-\frac{(\hat{z}_{1}-\hat{z}_{2})^{2}}{2}}}{\sqrt{2\pi}}
        \qty(-\mathcal{F}_{\rm{int}}\qty(\Delta(\hat{z}_{1},\hat{z}_{2})))
\;\;\;
\right\} \nonumber\\
\label{eq-F-large-d-glass-by-varphi-cavity-main-text}
\eeqn
where $\int_{\rm{ex-cav}} d\hat{z}$ is the integral outside the cavity
\begin{align}
    \int_{\rm{ex-cav}} d\hat{z}
    =\int^{\infty}_{-\infty} d\hat{z}
    -\int^{\hat{L}_{\rm{cav}}}_{0} d\hat{z}.
\end{align}

Then with the free energy given above,
we obtain the self-consistent equation for the order parameter in the cavity as
\begin{align}
    \label{self_consistent_equation_in_inhomogeneous_theory}
    \frac{1}{\Delta(\hat{z})}
    =\frac{\hat{\varphi}}{2}\int^{\infty}_{-\infty} \frac{d\hat{z}'}{\sqrt{2\pi}}
    e^{-\frac{(\hat{z}-\hat{z}')^2}{2}}
    \int_{-\infty}^{\infty} d\xi e^{\xi-\frac{1}{2}\Delta(\hat{z},\hat{z}')}g^{m}(\xi,\Delta(\hat{z},\hat{z}')) {f'}^2(\xi,\Delta(\hat{z},\hat{z}')) \qquad
     0 < \hat{z} < \hat{L}_{\rm cav},
\end{align}
This equation must be solved under the condition that $\Delta(\hat{z})=0$
outside the cavity, i.~e $\hat{z}<0$ or $\hat{L}_{\rm cav}<\hat{z}$
(see  \eq{eq-SP-HS-1RSB-cavity}).
Since we will consider volume fractions lower than that of the Kauzmann transition, we fix the
parameter $m$ as $m=1$ \cite{1130848328245474560} in the following.

\subsubsection{Hard-spheres}

We specifically analyzed the case of hard-spheres.
\blue{The technical details are reported in appendix \ref{appendix-HS}.}
The function $g(\xi,\Delta)$ which appear in
$\mathcal{F}_{\rm int}(\Delta)$ defined in \eq{eq-F-int-1RSB}
is obtained for the hard-spheres as,
\beq
g(\xi,\Delta)=e^{\frac{1}{2}\Delta\partial_{\xi}^{2}}\theta(\xi)=\Theta(\xi/\sqrt{2\Delta})
\eeq
where $\Theta(x)=\qty(1+\erf(x))/2$ and $\erf(x)$ is the error function.

First we analyzed the bulk (uniform) system
solving the saddle point equation \eq{self_consistent_eq_most_general}
with $\hat\varphi(\hat{z})=\hat\varphi$. Looking for the density at which the saddle point equation
(with $m=1$) disappears, we obtain the dynamical transition density of the
bulk system as
\beq
\hat\varphi_{\rm d}=4.8067787037 
\eeq

\begin{figure}[h]
    \includegraphics[width=0.9\textwidth]{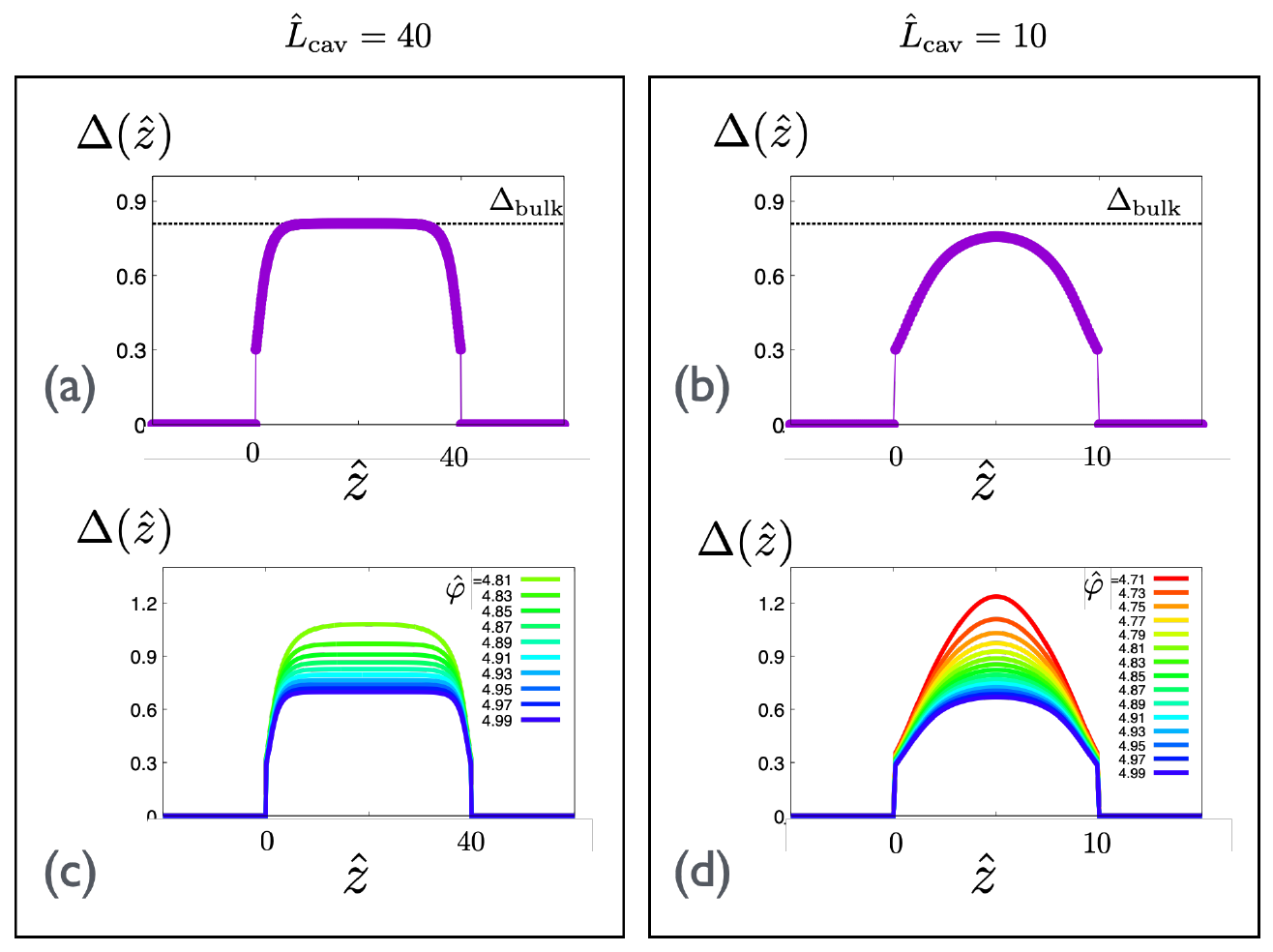}
    \caption{The spatial profile of the glass order parameter of the hard-spheres in the cavity: $\Delta(\hat{z})$ of $\hat{L}_{\rm cav}=10,40$
      at $\hat\varphi=4.90$ in (a),(b) and various other volume fractions $\hat\varphi$  (c),(d).
The dotted line in (a),(b) represents the order parameter $\Delta=\Delta_{\rm bulk}$ in bulk system ($\hat{L}_{\rm cav}=\infty$).
    }
    \label{fig-results-msd}
\end{figure}

We solve the saddle point equation numerically. \blue{For the details of the numerical procecure
  see appendix \ref{sec-numerical-detail-how-to-solve-SP}.}
In Fig.\ref{fig-results-msd} panel (a),(b) we show a representative spatial profile of the glass order parameter of the hard-spheres
in the cavity. The dotted lines represent $\Delta=\Delta_{\rm bulk}$ obtained in the bulk system ($\hat{L}_{\rm cav}=\infty$).
We can see that by moving far from the edge of the cavity the value of the order parameter 
becomes close to that of the bulk system.
Closer to the edge of the cavity, the order parameter becomes smaller meaning that the system
is more constrained there due to the frozen region outside the cavity.

\subsubsection{Cavity size dependence of the dynamical transition}

In Fig.\ref{fig-results-msd} panel (c),(d) we show the variation of the order parameter with respect to
the changes of the volume fraction $\hat\varphi$
of the hard-spheres in the cavity.
In the case $\hat{L}_{\rm cav}=40$
we found the solution disappears at $\hat\varphi\sim 4.8$ while the solution disappears
at $\hat\varphi\sim 4.7$ for $\hat{L}_{\rm cav}=10$. This suggest the dynamical transition
density $\hat\varphi_{\rm d}$ depends on the cavity size $\hat{L}_{\rm cav}$ such that
the $\hat\varphi_{\rm d}(\hat{L}_{\rm cav})$ becomes smaller decreasing $\hat{L}_{\rm cav}$.
This means the system in a smaller cavity is more strongly constrained due to the frozen region
outside the cavity  so that the glass state remains up to lower volume fractions.

\begin{figure}[h]
    \includegraphics[width=0.95\textwidth]{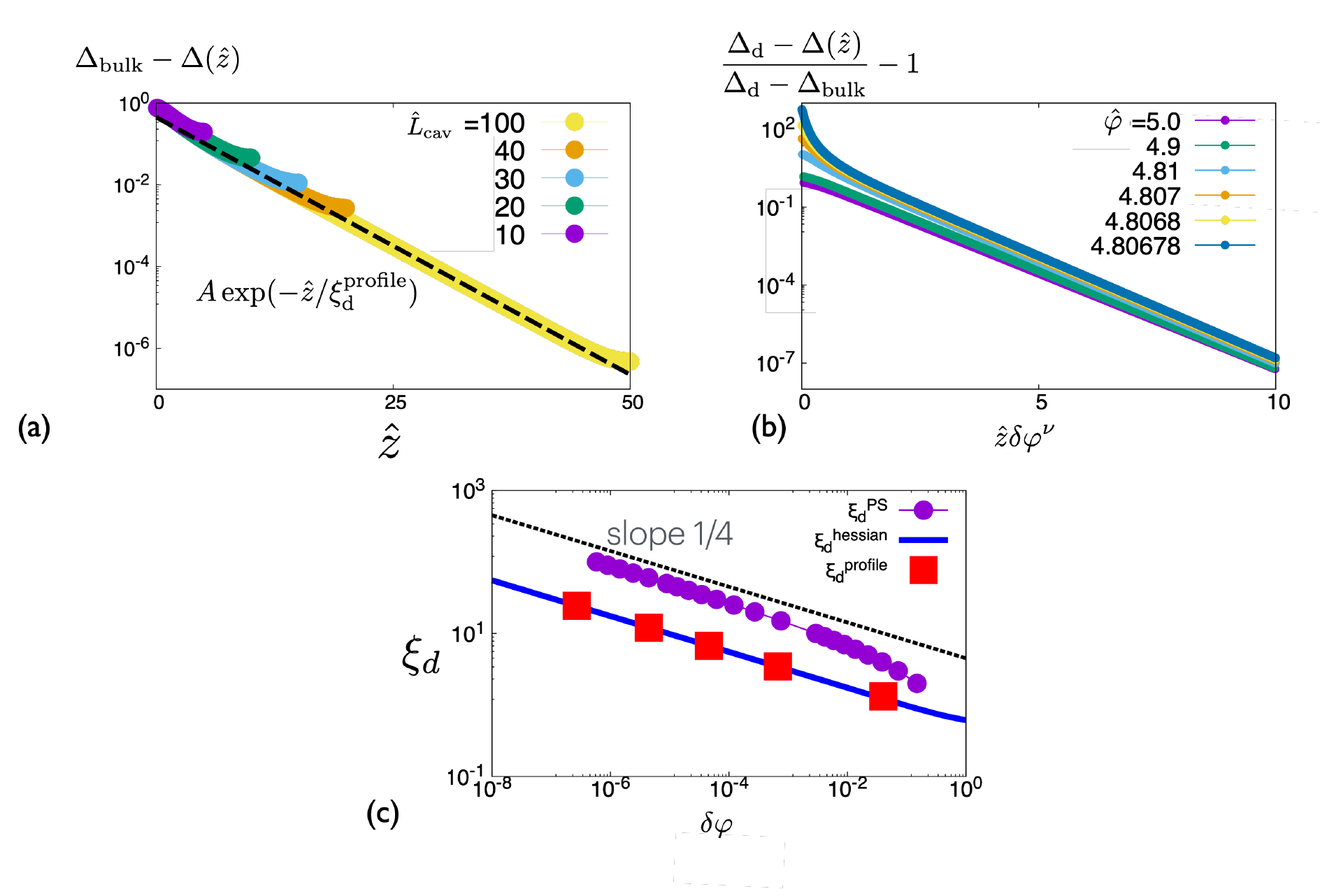}
    \caption{Scaling properties close to the dynamical transition density of the hard-spheres:
      Panel (a) shows the spatial profile of $\Delta_{\rm bulk}-\Delta(\hat{z},\delta\varphi))$ for various cavity sizes
      $\hat{L}_{\rm cav}=10,20,30,40$ at $\hat\varphi=4.81$
      (or $\delta\varphi=(\hat\varphi-\hat\varphi_{\rm d})/\hat\varphi_{\rm d}=6.70\times 10^{-4}$).
      The exponential fitting function
      \eq{eq-fit-exponential-delta-Delta} is also shown. Here
      we find $A=0.448192$ and $\xi_{\rm d}^{\rm profile}=3.44808$.
      Panel (b) shows a scaling plot of $\delta\Delta(\hat{z},\delta\varphi)$ \blue{with $\nu=1/4$
      using data of $\hat{L}_{\rm cav}=2000$.}
      Here $\delta\varphi\equiv(\varphi-\varphi_{d})/\varphi_{d}$
which represents the distance to the critical point.
      Panel c) displays various lengths diverging at the dynamical transition point:
      the PS length $\xi^{\rm PS}_{\rm d}$ (purple circles),
      the correlation length obtained
      from the spatial profile of $\delta \Delta(\hat{z},\delta\varphi)$ (see (a))
      (red dots) and the correlation length extracted in the analysis of the Hessian (blue line)( see \eq{eq-correlation-length-from-Hessian})
      vs $\delta\varphi$.
    }
        \label{fig-results-msd-scaling}
\end{figure}

We obtained the dynamical transition density $\hat{\varphi}_{d}(\hat{L}_{\rm cav})$, where the solution disappears,
for various cavity sizes $\hat{L}_{\rm cav}$. From this result we defined the point-to-set (PS) length
\cite{biroli2008thermodynamic}\cite{bouchaud2004adam}
of the dynamical transition $\xi^{\rm PS}_{d}(\hat\varphi)$ as
\beq
\xi^{\rm PS}_{d}(\hat\varphi_{\rm d}(\hat{L}_{\rm cav})) = \hat{L}_{\rm cav}/2
\eeq
treating $\hat{L}_{\rm cav}$ as a running parameter.
As shown in Fig.\ref{fig-results-msd-scaling} panel (c), $\xi^{\rm PS}_{d}(\hat\varphi)$ obeys well
the anticipated scaling
\beq
\xi^{\rm PS}_{\rm d} \propto  \xi^{\rm hessian}_{\rm d} \propto \delta \hat\varphi^{-1/4}.
\eeq

\subsubsection{Spatial profile of the glass order parameter around the dynamical glass transition}

We have seen that by moving far from the edge of the cavity the value of the order parameter $\hat\Delta(\hat{z})$
becomes close to that of the bulk system. To characterize such a spatial profile of the order parameter let us introduce
\beq
\delta \Delta(\hat{z})=\Delta_{\rm bulk}-\Delta(\hat{z})
\eeq
where $\Delta_{\rm bulk}$ is the value of the order parameter
of the bulk system $\hat{L}_{\rm cav}=\infty$.
As shown in Fig.~\ref{fig-results-msd-scaling} (a) it exhibits an exponential decay as a function of $\hat{z}$
, \blue{by choosing a large enough $\hat{L}_{\rm cav}$}, such that it can be well fitted by
\beq
\delta \Delta(\hat{z})
= A\exp \left (-\frac{\hat{z}}{\xi_{\rm d}^{\rm profile}}\right)
\label{eq-fit-exponential-delta-Delta}
\eeq
with a characteristic length scale $\xi_{\rm d}^{\rm profile}$.

In Fig.~\ref{fig-results-msd-scaling} (a) we find clear $\hat{L}_{\rm cav}$ dependence. In smaller $\hat{L}_{\rm cav}$ systems, the exponential decay saturates at smaller $\hat{z}$. It is natural to expect that the saturation will disappear in the
limit $\hat{L}_{\rm cav}/\xi_{\rm d}^{\rm profile} \to \infty$.
The red dots in Fig.~\ref{fig-results-msd-scaling} (c)
are $\xi_{\rm d}^{\rm profile}$ obtained analyzing the system with $\hat{L}_{\rm cav}=2000$.
\blue{We find $\xi_{\rm d}^{\rm profile}$
  clearly grows approaching the dynamical transition density $\hat\varphi_{\rm d}$
  similarly to $\xi_{\rm d}^{\rm hessian}$ and $\xi_{\rm d}^{\rm PS}$.}


Close to the critical point $\delta\varphi  \sim 0$,
it is natural to expect that the spatial profile of the order parameter, namely
$\Delta(\hat{z},\delta \varphi)$ as a function of $\hat{z}$ exhibits a universal scaling feature.
To see this, it is convenient to use the following decomposition,
\begin{align}
  \delta \Delta(\hat{z},\delta\varphi)=\Delta_{\rm bulk}(\delta\varphi)-\Delta(\hat{z},\delta\varphi)
  & =\Delta_{\rm d}-\Delta(\hat{z},\delta\varphi)+\Delta_{\rm bulk}(\delta\varphi)-\Delta_{\rm d} \nonumber \\
  & =\left(\Delta_{\rm d}-\Delta_{\rm bulk}(\delta\varphi)\right)f(\hat{z},\delta\varphi)
  \end{align}
where we introduced a dimensionless function,
\beq
f(\hat{z},\delta\varphi)
=
\blue{\frac{\Delta_{\rm d}-\Delta(\hat{z},\delta\varphi)}{\Delta_{\rm d}-\Delta_{\rm bulk}(\delta\varphi)}-1}
\eeq
where $\Delta_{\rm d}-\Delta_{\rm bulk} \sim \delta\varphi^{1/2}$ close to the critical point (see \eq{eq-order-parameter-scaling}). It is natural to expect that
$f(\hat{z},\delta\varphi)$ becomes a universal function
of $\hat{z}/\xi_{\rm d}^{\rm profile}$.
Note also that by definition
\beq
\blue{\lim_{\hat{z} \to \infty}} f(\hat{z},\delta\varphi)=0
\eeq
In fact Fig.\ref{fig-results-msd-scaling} (c) implies the scaling holds in $\delta\varphi \to 0$ limit.

\blue{One can expect that $\xi_{\rm d}^{\rm profile}$ agrees with $\xi_{\rm d}^{\rm hessian}$ based on the following argument.
In our setup the region outside the cavity consists of frozen, immobile spheres. Let us imagine
to replace them with spheres confined around the equilibrium reference points by some Hookian springs.
Indeed the field $\epsilon(\hat{z})$ which is introduced through \eq{eq-identity-for-q}
as the field conjugated to the order parameter $q_{ab}(\hat{z})$ and thus $\Delta(\hat{z})$
(see \eq{eq-def-alpha_ab} and  \eq{eq-def-Delta-alpha}) exactly serves as such a Hookian spring.
At the level of linear response, the change of the density profile $\delta \Delta(\hat{z})$
due to a change in the strength of the Hookian spring $\delta \epsilon(\hat{z}')$ is related to the spatial correlation function
of the spontaneous fluctuation of the mean-squared displacement $\delta \Delta(\hat{z})$ as,
\beq
\frac{\delta \Delta(\hat{z})}{\delta \epsilon(\hat{z}')}= \langle \delta\Delta(\hat{z})\delta\Delta(\hat{z}') \rangle
\eeq
This implies introduction of  Hookian springs of strength $\epsilon$ in the region $ -\infty < \hat{z}' < 0$
on an initially  homogeneous system amounts to inducing a change of the density profile at distance $\hat{z}$ from the wall as,
\beq
\delta \Delta (\hat{z})
=
\epsilon
\int_{-\infty}^{0}d\hat{z}'
\langle \delta\Delta(\hat{z})\delta\Delta(\hat{z}') \propto \epsilon\exp\left(
-\frac{\hat{z}}{\xi_{\rm d}^{\rm hessian}}
\right)
\eeq
where we used \eq{eq-correlation-function-of-Delta}. To recover the original setup in which the system is completely frozen outside the cavity we have to consider $\epsilon \to \infty$ which goes beyond  the simple linear-response argument.
Nonetheless the above argument hints the relation  $\xi_{\rm d}^{\rm profile}=\xi_{\rm d}^{\rm hessian}$.}

\subsubsection{Correlation length extracted from the fluctuation around the saddle point in the bulk system}

As we discussed in sec~\ref{sec-spatial-correlation-of-glassy-fluctuation-around-dynamical-transition}, characteristic
length scale associated with the dynamical transition can also be obtained by analyzing fluctuations around the saddle point.
For the hard spheres we obtained the key parameters
\beq
M_{0}=a\sqrt{\delta\varphi} \qquad a \simeq 0.635
\qquad M_{2} \simeq 0.376
\eeq
as explained in Appendix \ref{sec-longtidudinal-mode-of-hessian}.
In Fig.~\ref{fig-results-msd-scaling} (c) we also display  $\xi_{\rm d}^{\rm hessian}$ we obtained using $M_{0}$ and $M_{2}$ \blue{in \eq{eq-correlation-length-from-Hessian}}.
Remarkably $\xi_{\rm d}^{\rm hessian}$ perfectly matches with the characteristic length $\xi_{\rm d}^{\rm profile}$.

\subsubsection{Summary of the behavior of the system close to the dynamical transition point}

To summarize, we found three length scales i) the point-to-set length $\xi_{\rm d}^{\rm PS}$, ii) the correlation length of thermal fluctuation
$\xi_{\rm d}^{\rm hessian}$ and iii) the characteristic length of the spatial profile of the glass order parameter in the cavity
$\xi_{\rm d}^{\rm profile}$ all scales as $\delta\varphi^{-1/4}$ approaching the dynamical glass transition point $\hat\varphi_{\rm d}$.
Our result confirms that the point-to-set length is proportional to the correlation length. 

\blue{In the present paper we only considered the intra-state thermal fluctuation 
reflected in the longitudinal mode of the Hessian matrix for simplicity.
More generally one has to consider inter-state and sample-to-sample fluctuations
reflected in the replicon and anomalous modes all of which contribute
the so called $\chi_{4}$ \cite{franz2013static,franz2011field,PhysRevE.106.024605}.}

As far as we are aware of, the fact that the spatial profile of the glass order parameter $\Delta(\hat{z})$ also reflect a correlation length
$\xi_{\rm d}^{\rm profile}$ has not been appreciated in the context of glass physics (see however \cite{berthier2012static}).
As shown in  Fig.\ref{fig-results-msd-scaling} (c), we find $\xi_{\rm d}^{\rm profile}=\xi_{\rm d}^{\rm hessian}$.
The situation appears very similar to the surface critical phenomena, for example of ferromagnets,
where one observes that the spatial profile of the order parameter reflects the spatial correlation length of spontaneous thermal fluctuations
\cite{binder1972phase,kikuchi1985monte1,okabe1986general}.

\subsubsection{Cavity size dependence of the Kauzmann transition}

Finally let us focus on the Kauzmann transition in the cavity system.
The complexity (structural entropy) $\Sigma$ can be computed using the free-energy
\eq{eq-F-large-d-glass-by-varphi-cavity-main-text}
as explained in appendix~\ref{sec-kauzmann-transition-in-cavity}.

We find
\begin{align}
    \Sigma
    =\frac{d}{2}\qty[\ln{d}-\hat{\varphi}\qty(2-f(\hat{L}_{\rm cav}))].
\end{align}
with
\beq
f(L)=1-\sqrt{\frac{2}{\pi}}\hat{L}_{\rm{cav}}^{-1}+O\qty(e^{-\hat{L}_{\rm{cav}}^{2}/2}).
\eeq

The Kauzmann transition density of the cavity system obtained at $\Sigma=0$ is 
\begin{align}
    \label{cavity_kauzmann_eq}
    \hat{\varphi}_{K}(\hat{L}_{\rm cav})
    =\frac{\hat{\varphi}_{K,\rm{bulk}}}{2-f(\hat{L}_{\rm cav})}=
    \hat{\varphi}_{K,\rm{bulk}}\left(1+\sqrt{\frac{2}{\pi}}\hat{L}_{\rm cav}^{-1}+O\qty(e^{-\hat{L}_{\rm{cav}}^{2}/2}) \right)
\end{align}
where $\hat{\varphi}_{K,\rm{bulk}}$ is the Kauzmann transition density for the bulk system $\hat{L}_{\rm cav}=\infty$.
Thus in cavity systems the Kauzmann transition occurs at lower densities than in bulk systems and
the transition density increases increasing the cavity size $\hat{L}_{\rm cav}$.
Physically this can be understood again as the consequence of the constraint
imposed by the frozen system outside the cavity.

From this result we defined the point-to-set (PS) length
of the static transition $\xi^{\rm PS}_{\rm s}(\hat\varphi)$ as
\beq
\xi^{\rm PS}_{\rm s}(\hat\varphi_{\rm K}(\hat{L}_{\rm cav})) = \hat{L}_{\rm cav}/2
\eeq
treating $\hat{L}_{\rm cav}$ as a running parameter.
We find
\beq
\xi^{\rm PS}_{\rm s} \propto (\hat\varphi-\hat\varphi_{\rm K,bulk})^{-1}
\eeq

\section{Conclusion and outlook}

To conclude we constructed a framework of the inhomogeneous replicated liquid theory that can treat glasses
whose physical properties evolve along a one-dimensional axis. The theory becomes exact in the limit of infinite transverse dimensions.
Extension of the theory to describe $2$, $3$ dimensional inhomogeneities, \blue{with $\infty$ dimensional transverse space},
is straight forward. \blue{The dynamical version of our theory which extends the dynamical mean-field theory for the bulk system
\cite{maimbourg2016solution} can also be constructed straightfowradly which we will report elsewhere.
} 

In the present paper we successfully applied our scheme to analyze diverging length scales at dynamic/static glass transitions.
\blue{In particular we implemented a cavity setup to measure the point-to-set length
\cite{biroli2008thermodynamic}\cite{bouchaud2004adam} in the replicated liquid theory.
Our results reconfirm the critical exponents obtained in previous theoretical studies.}
\blue{Relevance of the static length scales in glassy dynamics may be questioned as it remains very small
when the dynamics become extremely slow \cite{PhysRevX.12.041028}. In the real glassy dynamics
other features such as avalanche like process which facilitates the relaxation and enhance the dynamical
heterogeneity may play important roles. However a recent work report that
in deeply annealed glass the point-to-set length
reasonably describes a dynamically observed characteristic length \cite{shiraishi2024characterizing}.}

\blue{The obvious limitation of our theory is that it is still very strongly 'mean-field' and miss many subtle features in real finite dimensional liquids.
  In particular the presence/absence of the 'dynamic glass transition' 
(or even the ideal 'static glass transition')
in realistic systems cannot be addressed.
 However we believe it will be useful as a tool to study properties of the solid, glassy states
at low temperatures/ high densities, which have sufficiently long life-times,  from 1st principles. 
It is straightforward to adapt the glass state following \cite{rainone2015following} scheme
in our setup. It will be particularly interesting to apply such a scheme to study
the emergence of spatial inhomogeneities in amorphous solids under compression, shear  e.t.c.
approaching yielding and jamming \cite{jin2017exploring,jin2018stability,jin2021jamming,pan2023review}.
Both yielding and jamming require precise microscopic descriptions
as demonstrated by the $d\to \infty$ theories \cite{charbonneau2014exact,rainone2015following}.
However the characteristic spatial heterogeneities \cite{wyart2005rigidity,maloney2006amorphous} associated with them have not been studid by 1st principle theories.
Our $1+\infty$ dimensional
theory will be a promising tool to investigate these problems. 
}

\begin{acknowledgements}
  We thank Yuki Rea Hamano, 
  Yusuke Hara, Atsushi Ikeda, Harkuni Ikeda, Macoto Kikuchi, Hyonggi Kim, Kang Kim, Norifumi Maruyama, Kota Mitsumoto
  Hideyuki Mizuno, Misaki Ozawa, Yuki Takaha, Touta Yoshida,
  Akira Warita and Francesco Zamponi for useful discussions. This work was supported by KAKENHI
  (No. 21K18146) (No. 20H00128) from MEXT, Japan  and  JST SPRING, Grant Number JPMJSP2138.
\end{acknowledgements}

\newpage
\appendix
\section{Liquid theory in $1+(d-1)$ dimensions with $d-1 \gg 1$}
\label{appendix-liquid-theory}

\subsection{Basic setup}
\label{appendix-liquid-theory-basic-setup}

Here we develop a density functional expression of the free-energy of the system of $N$ particles given by the Hamiltonian \eq{hamiltonian_of_liquid_theory}.
Let us write the number density (per unit length), which varies along $z$ as,
\beq
\hat{\rho}_{\rm micro}(z)=\sum_{i=1}^{N} \delta(z-z_{i}),
\eeq
and introduce an identity,
\beq
1=\int {\cal D}\hat{\rho}(z) \delta (\hat{\rho}(z)-\hat{\rho}_{\rm micro}(z))
=\int {\cal D} \hat{\rho}(z){\cal D} \phi(z) \exp \left [\int dz \phi(z)(\hat{\rho}(z)-\hat{\rho}_{\rm micro}(z)) \right]
\label{eq-identity}
\eeq
where ${\cal D}\hat{\rho}(z)$ is a functional integration over $\hat{\rho}(z)$ and $\delta(...)$ is a functional delta.
In the 2nd equation we introduced an integral representation of the functional delta introducing a function $\phi(z)$
which can be related to the so called intrinsic chemical potential \cite{hansen1990theory}.

Then the free-energy of the system  can be written as
\beq
-\beta F=\ln Z \qquad 
Z=\frac{1}{N!}\int \prod_{i=1}^{N} \frac{d^{d-1}x_{i}}{\lambda_{\rm th}^{d-1}}\frac{dz_{i}}{\lambda_{\rm th}}
e^{-\beta \sum_{i<j} v(r_{ij})-\beta \sum_{i=1}^{N}U(z_{i})}
\eeq
\blue{Inserting '1' in the last equation, using the identity \eq{eq-identity},
and performing some interchanges of the order of integrals 
one finds that the  free-energy $F$ can be expressed formally as,}
\beq
e^{-\beta F}=\int {\cal D} \hat{\rho}(z) e^{-\beta F[\hat{\rho}]}
\label{eq-F-total}
\eeq
with a functional $F[\hat{\rho}]$ defined as,
\beq
e^{-\beta F[\hat{\rho}]}=\int {\cal D} \phi(z) e^{\int dz \phi(z)\hat{\rho}(z)}
e^{-\beta G[\phi]} \label{eq-laplace}
\eeq
Here the functional $G[\phi]$ is defined as,
\beqn
e^{-\beta G[\phi]}&=&
\frac{1}{N!}\int \prod_{i=1}^{N}d^{d-1}x_{i} dz_{i}
\prod_{i=1}^{N}a(z_{i}) 
\prod_{i< j}(1+ \lambda f(r_{ij}))
\label{eq-G}
\eeqn
In \eq{eq-G} we also introduced the Mayer function\cite{hansen1990theory},
\beq
f(r)=e^{-\beta v(r)}-1
\label{eq-mayer}
\eeq
and the 'activity',
\beq
a(z)=\frac{e^{-\phi(z)-\beta U(z)}}{\lambda_{\rm th}^{d}}
\eeq
In \eq{eq-G} we also introduced a parameter $\lambda$ to organize the Mayer expansion discussed below. It will be put back to $1$ after organizing the expansion.

\blue{Physically the functionals $F[\rho]$ and $G[\phi]$ introduced above can be regarded as a free-energy functionals.}
We anticipate that the functional integration in \eq{eq-laplace} can be done by the saddle point method for $N \gg 1$ which yields,
\beq
-\beta F[\hat{\rho}]=\int dz \phi^{*}(z)\hat{\rho}(z) - \beta G[\phi^{*}]
\label{eq-legendre}
\eeq
with the saddle point $\phi^{*}(z)=\phi^{*}[\hat{\rho}](z)$ is determined by
\beq
\hat{\rho}(z)=\left. \frac{\delta (-\beta G[\phi])}{\delta(-\phi(z))} \right|_{\phi=\phi^{*}[\hat{\rho}]}
\label{eq-rho-phi}
\eeq
\blue{Thus \eq{eq-legendre} is a Legendre transformation associated with the Laplace transformation 
\eq{eq-laplace}.}

Now we evaluate $G[\phi]$ treating the effect of interactions perturbatively, i.~e. Mayer expansion.
\beq
e^{-\beta G[\phi]}= e^{-\beta G_{0}[\phi]}[1+ \lambda \sum_{i< j} \langle f(r_{ij}) \rangle_{\phi} + O(\lambda^{2})]
\label{eq-G-expansion}
\eeq
Here $-\beta G_{0}[\phi]$ is the free-energy of the ideal gas which is obtained using the Stirling's formula $\ln N!\sim N\ln N -N$,
\beq
-\beta G_{0}[\phi]=-N \ln N + N + N \ln \left [S\int dz a(z) \right]
\label{eq-G0}
\eeq
with $S=\int d^{d-1}x$ being the surface area of the cross-section (See Fig.~\ref{fig_cylinder}).
We also introduced
\beq
\langle \ldots \rangle_{\phi} =\frac{\prod_{i=1}^{N} \int d^{d-1}x_{i}dz_{i} a(z_{i}) \ldots}{\prod_{i=1}^{N} \int d^{d-1}x_{i}dz_{i} a(z_{i})}
\eeq
which is the thermal average took within the non-interacting system.

The outline of the analysis goes as follows. We consider series expansions,
\beqn
\phi^{*} = \phi^{*}_{0}+\lambda \phi^{*}_{1} + \frac{\lambda^{2}}{2} \phi^{*}_{2} + \ldots \qquad
G  = G_{0}+\lambda G_{1} + \frac{\lambda^{2}}{2} G_{2} + \ldots \qquad
F  = F_{0}+\lambda F_{1} + \frac{\lambda^{2}}{2} F_{2} + \ldots
\label{eq-legendre-transform-expansion-plefka}
\eeqn
The series for $G$ can be determined by analyzing \eq{eq-G-expansion}. Using the result in 
\eq{eq-rho-phi} the series for $\phi^{*}$ can be obtained. Finally using these in
\eq{eq-legendre-transform-expansion-plefka}
we will obtain the series for $F$.  Up to 1st order one can find,
\beqn
-\beta F_{0}[\hat{\rho}] &=&\int dz \phi_{0}^{*}[\hat{\rho}](z)\hat{\rho}(z)-\beta G_{0}[\phi^{*}_{0}[\hat{\rho}]] \label{eq-F0}\\
-\beta F_{1}[\hat{\rho}] &=& -\beta G_{1}[\phi^{*}_{0}[\hat{\rho}]] \label{eq-F1}
\eeqn
To work out $\phi_{0}^{*}$ explicitly we use \eq{eq-G0} in \eq{eq-rho-phi} and find,
\beq
\hat{\rho}(z)=N\frac{a^{*}(z)}{\int dz a^{*}(z)} \qquad a^{*}(z)=\frac{e^{-\phi^{*}(z)-\beta U(z)}}{\lambda_{\rm th}^{d}}.
\eeq
Using this in \eq{eq-F0} we find the ideal-gas part or the entropic part of the free-energy as,
\beq
-\beta F_{0}[\hat{\rho}]=\int dz \hat{\rho}(z)\left[1-\ln (\lambda_{\rm th}^{d}\frac{\hat{\rho}(z)}{S})\right]
+\int dz \hat{\rho}(z)(-\beta U(z))
\eeq

Now let us consider the effect of interactions. From \eq{eq-G-expansion} we find,
\beq
-\beta G_{1}[\phi]=\sum_{i < j} \langle f(r_{ij}) \rangle_{\phi} 
=\frac{N(N-1)}{2}
\frac{\prod_{i=1}^{2} \int d^{d-1}x_{i}dz_{i} a(z_{i}) f(r_{12})}{\prod_{i=1}^{2} \int d^{d-1}x_{i}dz_{i} a(z_{i})}
\eeq
Using this in \eq{eq-F1} we find,
\beq
-\beta F_{1}[\hat{\rho}]=\frac{1}{2}\int dz_{1} dz_{2} \hat{\rho}(z_{1})\hat{\rho}(z_{2})
\int \frac{d^{d-1}x_{1}}{S}\int \frac{d^{d-1}x_{2}}{S}f(r_{12})
\eeq

Collecting the above results we obtain up to 1st order in the Mayer-expansion,
\beqn
\frac{-\beta F[\rho]}{S}&=&\int dz \rho(z)\left[1-\ln (\lambda_{\rm th}^{d}\rho(z))\right]
+\int dz \rho(z) (-\beta U(z)) \nonumber \\
&+&\frac{1}{2}\int dz_{1} dz_{2} \rho(z_{1})\rho(z_{2})
\frac{1}{S}\int d^{d-1}x_{1}\int d^{d-1}x_{2}f(r_{12})
\label{eq-F}
\eeqn
where we introduced
\beq
\rho(z)=\frac{\hat{\rho}(z)}{S}.
\label{eq-rho-and-hat-rho}
\eeq
which is the number density field per unit volume.
Note that $\rho(z)$  must be normalized such that,
\beq
S \int dz \rho(z)=N
\label{eq-normalization-rho}
\eeq
Finally the thermodynamic free-energy $F$ can be obtained through \eq{eq-F-total} where the functional integral
can be evaluated by the saddle point method for $N \gg 1$,
\beq
F=F[\rho^{*}] \qquad 0=\left. \frac{\delta}{\delta \rho(z)}
\left\{ \frac{-\beta F[\rho]}{S}+\beta\mu \left(\int  dz \rho(z)-\frac{N}{S}\right)
\right \} \right |_{\rho=\rho^{*}}
\label{eq-devF-via-rho}
\eeq
Here we introduced a Lagrange multiplier $\beta \mu$ to impose the normalization condition \eq{eq-normalization-rho}
where $\mu$ can be regarded as the chemical potential.

As is well known \cite{hansen1990theory}, the contributions from higher order terms in the Mayer expansion
into the free-energy $F[\rho]$ can be represented by one-particle irreducible diagrams.
It has been shown that they become negligible in the large dimensional limit $d \to \infty$
\cite{frisch1985classical,PhysRevA.36.2422,1130848328245474560}.

\subsection{Large dimensional limit}
\label{appendix-liquid-theory-large-dim-limit}

Here we derive an expression of the free-energy functional useful in $d-1 \gg 1$ limit.
We consider two-body potential $v(r)$ characterized by a microscopic length scale $D$
such that it becomes a function of $\xi$ defined as,
\beq
r_{\perp}=D\left (1+\frac{\xi}{d-1} \right)
\label{eq-xi}
\eeq
in $d \to \infty$ limit \cite{1130848328245474560}.
Then we can write
\beqn
\lim_{S \to \infty}\frac{1}{S}\int d^{d-1}x_{1} \int d^{d-1}x_{2}f(r_{12}) &=&
\Omega_{d-1}(d-1) \int_{0}^{\infty} dr_{12,\perp}r_{12,\perp}^{d-1} f(r_{12,\perp}^{2}+(z_{1}-z_{2})^{2}) \nonumber \\
&\xrightarrow{d\to \infty}&  \Omega_{d-1} D^{d-1}\int_{-\infty}^{\infty} d\xi e^{\xi} f\left[D^{2}\left(
1+\frac{1}{d}\left(\xi+
\frac{(\hat{z}_{1}-\hat{z}_{2})^{2}}{2}
\right)+O(d^{-2})
  \right)^{2}\;\;
  \right] \qquad
\eeqn
where $\Omega_{d}$ is the volume of $d$-dimensional unit sphere.  We also introduced a scaled coordinate $\hat{z}$ such that
\beq 
z=\frac{D}{\sqrt{d}}\hat{z}
\label{eq-z-and-scaled-z}
\eeq

Finally we obtain
\beqn
-\frac{\beta F[\rho]}{
  SD/\sqrt{d}} &=&\int d\hat{z}\rho(\hat{z})
\left [1-
  \ln \left (\lambda_{\rm th}^{d} \rho(\hat{z}) \right) \right]
+\int d\hat{z} \hat{\rho}(\hat{z}) (-\beta U(\hat{z}))
\nonumber \\
&+&\frac{d}{2} \frac{\Omega_{d}}{d} D^{d}
\int d\hat{z}_{1} \rho(\hat{z}_{1})
\int d\hat{z}_{2} \rho(\hat{z}_{2})
\frac{e^{-\frac{(\hat{z}_{1}-\hat{z}_{2})^{2}}{2}}}{\sqrt{2\pi}}
(-{\cal F})
\label{eq-F-large-d}
\eeqn
using $\Omega_{d-1}=\sqrt{d/2\pi}\Omega_{d}$. Here we introduced
\beq
-{\cal F}=
\int_{-\infty}^{\infty}d\xi e^{\xi}
f\left[D^{2}\left(
1+\frac{\xi}{d}
  \right)^{2}
  \right]
\label{eq-cal-F}
\eeq

The equilibrium density  profile $\rho^{*}(z)$ is obtained as \eq{eq-devF-via-rho}
using the chemical potential, 
\beqn
-\beta \mu  &=&
\left. \frac{\delta}{\delta \rho(\hat{z}_{1})}
\frac{-\beta F[\rho]}{
  SD/\sqrt{d}}
\right |_{\rho=\rho^{*}} \nonumber \\
&=&-\ln (\lambda_{\rm th}^{d}\rho^{*}(\hat{z}_{1}))
+ (-\beta U(\hat{z}_{1}))
+d \frac{\Omega_{d}}{d} D^{d}
\int d\hat{z}_{2} \rho^{*}(\hat{z}_{2})
\frac{e^{-\frac{(\hat{z}_{1}-\hat{z}_{2})^{2}}{2}}}{\sqrt{2\pi}}
(-{\cal F})
\label{eq-mu}
\eeqn
where the chemical potential $\mu$ should be chosen such that 
\beq
S \int dz\rho(\hat{z})=
\frac{SD}{\sqrt{d}}\int d\hat{z}\rho(\hat{z})=
N
\label{eq-normalization-rho-mod}
\eeq
becomes satisfied.

In the case of hard-sphere like systems, $D$ is the radius of particles. For those cases
it is convenient to introduce the volume fraction $\varphi(\hat{z})$,
\beq
    \varphi(\hat{z})
    =\rho(\hat{z})\Omega_{d}\qty(\frac{D}{2})^{d}
    \label{eq-var-phi}
\eeq
In order to study glassy sates in the large dimensional limit it is convenient
to introduce a scaled volume fraction \cite{1130848328245474560},
\beq
    \hat{\varphi}(\hat{z})
    \equiv\frac{2^{d}\varphi(\hat{z})}{d}
    =\frac{\rho(\hat{z})\Omega_{d} D^{d}}{d}.
    \label{eq-hat-var-phi}    
    \eeq
    Using this the free-energy can be expressed as
\beqn
-\frac{\beta F[\hat\varphi]}{
  SD/\sqrt{d}}
\frac{\Omega_{d}D^{d}}{d}
&=&\int d\hat{z}\hat{\varphi}(\hat{z})
\left [1-
  \ln \left (\hat{\varphi}(\hat{z}) \right)
+(-\beta \hat{U}_{0}(\hat{z}_{1}))
  \right]
\nonumber \\
&+& d \left\{
\int d\hat{z} \hat{\varphi}(\hat{z}) (-\beta \hat{U}_{1}(\hat{z}))
+\frac{1}{2}
\int d\hat{z}_{1} \hat{\varphi}(\hat{z}_{1})
\int d\hat{z}_{2} \hat{\varphi}(\hat{z}_{2})
\frac{e^{-\frac{(\hat{z}_{1}-\hat{z}_{2})^{2}}{2}}}{\sqrt{2\pi}}
(-{\cal F})
\right\} \qquad\qquad 
\eeqn
where we omitted a constant $N(\Omega_{d}D^{d}/d)\ln((\lambda_{\rm th}/D)^{d}(d/\Omega_{d}))$.
Here we also introduced a parametrization of the external potential,
\beq
U(\hat{z})=\hat{U}_{0}(\hat{z})+d\hat{U}_{1}(\hat{z})
\eeq
which is convenient to consider $d \to \infty$ limit.
Similarly the chemical potential
can be expressed as
\beqn
-\beta \mu
&=&-\ln((\lambda_{\rm th}/D)^{d}(d/\Omega_{d}))
-\ln (\hat\varphi^{*}(\hat{z}_{1}))+(-\beta \hat{U}_{0}(\hat{z}_{1}))
 \nonumber \\
&+& d \left \{
(-\beta \hat{U}_{1}(\hat{z}_{1}))
+\int d\hat{z}_{2} \hat{\varphi}^{*}(\hat{z}_{2})
\frac{e^{-\frac{(\hat{z}_{1}-\hat{z}_{2})^{2}}{2}}}{\sqrt{2\pi}}
(-{\cal F})
\right \}
\label{eq-chemical-potential-liquid}
\eeqn
This expression implies that the external potential can be designed
to realize any desired density profile $\hat{\varphi}(z)=O(1)$.

\subsection{Compression}
\label{sec-compression-liquid}

Let us discuss compression (or decompression) of our system.
To this end we parameterize the changes of the volume as
\beq
V(\eta)=V_{0}e^{-\eta}
\label{eq-volume-eta}
\eeq
Thus we are compressing for $\eta >0$ and decompressing for $\eta < 0$.
A change of the volume amounts to a change of the boundary condition.
By writing the original coordinate system as ${x'_{1},x'_{2},\ldots,x'_{d-1},z'}$,
we can introduce a new coordinate system ${x_{1},x_{2},\ldots,x_{d-1},z}$
with $x_{\mu}=x'_{\mu}(1+\eta/d)$ for $\mu=1,2,3,\ldots,d-1$ and $z_{\mu}=z'_{\mu}(1+\eta/d)$.
With the new coordinate system, the boundary condition is brought back to the original one.

Then the expression of the free-energy \eq{eq-F-total} becomes
\beqn
-\beta F(\eta) &=& \ln 
\frac{1}{N!}\int_{V(\eta)} \prod_{i=1}^{N} \frac{d^{d-1}(x_{i})'}{\lambda_{\rm th}^{d-1}}\frac{dz'_{i}}{\lambda_{\rm th}}
e^{-\beta \sum_{i<j} v(r'_{ij})-\beta \sum_{i=1}^{N}U(z'_{i})} \nonumber \\
&=& \ln 
\frac{1}{N!}
\underbrace{\left(1-\frac{\eta}{d}\right)^{d}}_{\mbox{$e^{-\eta}$ in $d\to\infty$}}
\int_{V(0)} \prod_{i=1}^{N} \frac{d^{d-1}x_{i}}{\lambda_{\rm th}^{d-1}}\frac{dz_{i}}{\lambda_{\rm th}}
e^{-\beta \sum_{i<j} v(r_{ij}(1-\eta/d))-\beta \sum_{i=1}^{N}U(z_{i}(1-\eta/d))}
\eeqn
Then the free-energy functional \eq{eq-F-large-d} becomes
\beqn
-\frac{\beta F[\rho,\eta]}{
  SD/\sqrt{d}} &=&\int d\hat{z}\rho(\hat{z})
\left [1-\eta-
  \ln \left (\lambda_{\rm th}^{d} \rho(\hat{z}) \right) \right]
+\int d\hat{z} \hat{\rho}(\hat{z}) (-\beta U(\hat{z}(1-\eta/d)))
\nonumber \\
&+&\frac{d}{2} \frac{\Omega_{d}}{d} D^{d}
\int d\hat{z}_{1} \rho(\hat{z}_{1})
\int d\hat{z}_{2} \rho(\hat{z}_{2})
\frac{e^{-\frac{(\hat{z}_{1}-\hat{z}_{2})^{2}}{2}}}{\sqrt{2\pi}}
\int_{-\infty}^{\infty}d\xi e^{\xi}
f\left[D^{2}\left(
1+\frac{\xi}{d}-\frac{\eta}{d}
  \right)^{2}
  \right]
\label{eq-F-large-d-eta}
\eeqn

We can compute the pressure as
\beq
P=-\frac{\partial F}{\partial V} =\frac{\partial}{\partial \eta}
\frac{F}{V}
\label{eq-pressure}
\eeq
using $V\frac{\partial}{\partial V}=-\frac{\partial}{\partial \eta}$.
The reduced pressure is obtained as,
\beqn
&&p=\frac{\beta P}{\rho}=\frac{\partial}{\partial \eta}\frac{\beta F}{N} 
=-\frac{\partial}{\partial \eta} \frac{-\beta F}{SD/\sqrt{d}\int d\hat{z}\rho(\hat(z))} \nonumber \\
&&= 1 +\frac{1}{d}
\frac{\int d\hat{z} \hat{\rho}(\hat{z})(-\beta U'(\hat{z}(1-\eta/d)))
  }{\int d\hat{z} \hat{\rho}(\hat{z})}
 \nonumber \\
&+&\left[\int d\hat{z} \hat{\rho}(\hat{z})\right]^{-1}
\left[
  \frac{d}{2} \frac{\Omega_{d}}{d} D^{d}
\int d\hat{z}_{1} \rho(\hat{z}_{1})
\int d\hat{z}_{2} \rho(\hat{z}_{2})
\frac{e^{-\frac{(\hat{z}_{1}-\hat{z}_{2})^{2}}{2}}}{\sqrt{2\pi}}
\underbrace{\left (-\frac{\partial}{\partial \eta} \right)
\int_{-\infty}^{\infty}d\xi e^{\xi}
f\left[D^{2}\left(
1+\frac{\xi}{d}-\frac{\eta}{d}
\right)^{2}\right]}_{-\int_{-\infty}^{\infty}d\xi e^{\xi}
f\left[D^{2}\left(
1+\frac{\xi}{d}-\frac{\eta}{d}
\right)^{2}\right]}
\right]
\label{eq-reduced-pressure-liquid-theory}
\eeqn
with $\rho=N/V$.
We have $N=(SD/\sqrt{d})\int d\hat{z} \rho(\hat{z})$
(see \eq{eq-normalization-rho-mod}). In the last equation we performed
an integration by parts.

In the 2nd term of \eq{eq-reduced-pressure-liquid-theory} we find
a contribution to the pressure due to the external potential. Disregarding
the latter, the reduced pressure verifies
\beq
p\int d\hat{z}\rho(\hat{z})= \int d\hat{z} \rho(\hat{z}) \beta \mu -\frac{\beta F}{SD/\sqrt{d}}
\eeq
which is equivalent to the thermodynamic relation $PV=\mu N-F$.

\section{Replicated liquid theory in $1+(d-1)$ dimensions with $d-1 \gg 1$}
\label{appendix-replicated-liquid-theory}

\subsection{Basic setup}
\label{sec-basic-setup-replicated-liquid-theory}

 Now we turn to derive the free-energy expression for the glassy states. The free-energy $F$ of the system
 is related to the logarithm of the partition function which can be expressed in terms of a replicated system,
\beq
-\beta F=\ln Z \qquad  \ln Z = \left. \partial_{n} Z^{n}\right|_{n=0}
\eeq
where the partition function of the replicated systems $a=1,2,\ldots,n$ is given by,
\beqn
Z^{n}&=&\frac{1}{(N!)^{n}}\prod_{a=1}^{n}\int \prod_{i=1}^{N} \frac{d^{d-1}x^{a}_{i}}{\lambda_{\rm th}^{d-1}}\frac{dz^{a}_{i}}{\lambda_{\rm th}}e^{-\beta \sum_{i< j}v(r_{ij}^{a})-\beta \sum_{i} U(z_{i}^{a})} 
=\prod_{{\cal C}=1}^{n/m} Z_{m}
\eeqn
In the last equation, anticipating the spontaneous formation of the molecular liquid state,
we divided the $n$ replicas
into $n/m$ subgroups ${\cal C}=1,2,\ldots, n/m$ each of
which consists of $m$ replicas. Group ${\cal C}=1$ consists of replicas $a=1,2,\ldots,m$,
${\cal C}=2$ consists of replicas $a=m+1,m+2,\ldots,2m$ and so on.
Thus  we can write,
\beq
-\beta F=\frac{1}{m} \log Z_{m}
\label{eq-F-total-glass}
\eeq

It has been established in $d\to \infty$ \cite{1130848328245474560} that
the parameter $m$ should be set as follows. In the genuine liquid phase
at high enough temperatures/small enough densities, $m=1$.
At low enough temperatures/large enough densities beyond the so called Kauzmann transition
(static glass transition) 
ideal glass phase can emerge where $m$ should be chosen such that the complexity remains $0$ \cite{Mo95}.
There is an intriguing intermediate temperature/density
bounded by the so called dynamical glass transition and the Kauzmann transition.
There are a large number of glassy metastable states emerge 
but the system is categorized still as a liquid $m=1$ in the thermodynamic sense.
The information of the glassy metastable states is contained
in the so called Franz-Parisi's potential \cite{FP95} which is
a term proportional to $1-m$ within the replicated free-energy $-\beta m F$
\eq{eq-F-total-glass}.

The partition function associated with the group of $m$ replicas reads as,
\beqn
Z_{m}&=&
\frac{1}{(N!)^{m}} \prod_{a \in {\cal C}}
\int \prod_{i=1}^{N} \frac{d^{d-1}x^{a}_{i}}{\lambda_{\rm th}^{d-1}}\frac{dz^{a}_{i}}{\lambda_{\rm th}}e^{-\beta \sum_{i< j}v(r_{ij}^{a})-\beta \sum_{i} U(z_{i}^{a})} \nonumber \\
&=&\frac{1}{N!}
\prod_{i=1}^{N}
\left(\frac{m}{\lambda_{\rm th}}\right)^{d}
\int d(z_{i})_{\rm c}\int d^{d-1}(x_{i})_{\rm c}
\prod_{\mu=1}^{d}
\left \{
\prod_{a \in {\cal C}}d(u_{i}^{\mu})^{a} \delta(\sum_{a \in {\cal C}} (u_{i}^{\mu})^{a})
\right \}
\prod_{a \in {\cal C}}
e^{-\beta \sum_{i< j}v(r_{ij}^{a})-\beta \sum_{i} U(z_{i}^{a})} 
\eeqn
In the last equation, again anticipating the spontaneous formation of the molecular liquid state,
we introduced new coordinates for spatial integrations
writing the coordinate  $\bm{x}_{i}^{a}$ of particle $i$ in the $a$ replica as,
\begin{align}
    \bm{x}_{i}^{a}
    =(\bm{x}_{i})_{\rm c}+\bm{u}_{i}^{a}
    \label{eq-decomposition-coordinate}
\end{align}
This representation is convenient when the replicated liquid becomes a 'molecular liquid'
with 'molecules' $i=1,2,\ldots,N$  whose centers of mass are located at
\beq
(\bm{x}_{i})_{\rm c}=((x^{1}_{i})_{\rm c},(x^{2}_{i})_{\rm c},\ldots,(x^{d}_{i})_{\rm c}
,(z_{i})_{\rm c})=\frac{1}{m}\sum_{a=1}^{m}\bm{x}_{i}^{a} 
\eeq
The 2nd term in the r.h.s of \eq{eq-decomposition-coordinate}
represents the thermal fluctuations within the molecules.
Note that a sum rule
\beq
\sum_{a=1}^{m}\bm{u}_{i}^{a}=0
\label{eq-sum-rule}
\eeq
must hold for the relative coordinates. Note also that there are $(N!)^{m-1}$ different permutations of the particles labels by $i=1,2,\ldots,N$ and $a=1,2,\ldots,m$
to form such molecules and that the system (Hamiltonian) is invariant under these permutations.

In the following we will use the molecular coordinate \eq{eq-decomposition-coordinate}
{\it but drop the subscript $c$ for the center of mass to lighten the notations}.
Let us write the number density (per unit length) field of the 'molecules', which varies along $z$ as,
\beq
\hat{\rho}_{\rm micro}(z)=\sum_{i=1}^{N} \delta(z-(z_{i})_{\rm c}),
\eeq
and introduce an identity,
\beq
1=\int {\cal D}\hat{\rho}(z) \delta (\hat{\rho}(z)-\hat{\rho}_{\rm micro}(z))
=\int {\cal D} \hat{\rho}(z){\cal D} \phi(z) \exp \left [\int dz \phi(z)(\hat{\rho}(z)-\hat{\rho}_{\rm micro}(z)) \right]
\eeq
Let us introduce the space dependent glass order parameter $q_{ab}(z)$ as,
\beq
q_{ab}(z)\hat{\rho}(z)=\sum_{i=1}^{N} \left\langle ({\bm  u}_{i})^{a}\cdot ({\bm u}_{i})^{b}\delta(z-z_{i}) \right\rangle
\eeq
where $\langle \ldots \rangle$ is the appropriate thermal average.
Finiteness of it means formation of a molecular liquid state, i.~e. a glass state.
On the contrary, in a genuine liquid state, such molecules should be dissociated so that this parameter diverges.
Based on this observation let us introduce another identity,
\beqn
1&=&\int {\cal D}[q_{ab}(z)]\int {\cal D}[\epsilon_{ab}(z)]
\exp \left[
  -\frac{1}{2}\int dz \sum_{a,b=1}^{m-1}\epsilon_{ab}(z)
  \left[
    q_{ab}(z)\hat{\rho}(z)-\sum_{i=1}^{N}
        \sum_{\mu=1}^{d}    (u_{i}^{\mu})^{a}(u_{i}^{\mu})^{b}
    \right]
  \right]
\label{eq-identity-for-q}
\eeqn
Note that
\beq
\sum_{a=1}^{m}q_{ab}(z)=\sum_{b=1}^{m}q_{ab}(z)=0
\label{eq-sum-rule-q}
\eeq
because of the sum rule \eq{eq-sum-rule}. Thus we find
 $q_{am}=-\sum_{b=1}^{m-1}q_{ab}$, $q_{mb}=-\sum_{a=1}^{m-1}q_{ab}$
and $q_{mm}=\sum_{b=1}^{m-1}\sum_{b=1}^{m-1}q_{ab}$.

Using these we can write,
\beq
Z_{m}=\int {\cal D}\hat{\rho}(z)\int {\cal D}[q_{ab}(z)]e^{-\beta F_{m}[\hat{\rho}(z),q_{ab}(z)]}
\label{eq-partition-function-zm}
\eeq
where we defined
\beqn 
e^{-\beta F_{m}[\hat{\rho},q_{ab}]}=
\int {\cal D}\phi(z) e^{\int dz \phi(z)\hat{\rho}(z)}
\int {\cal D}[\epsilon_{ab}(z)] e^{-\frac{d}{2}\int dz \epsilon_{ab}(z)q_{ab}(z)\hat\rho(z)}
e^{-\beta  G_{m}[\phi,\epsilon_{ab}]}
\label{eq-legendre-replica}
\eeqn
with
\beqn
e^{-\beta  G_{m}[\phi,\epsilon_{ab}]}
&=&e^{-\beta  G_{0,m}[\phi,\epsilon_{ab}]}
\left[1+\lambda \sum_{i<j} \langle f_{m}(\{r^{a}_{ij}\}) \rangle_{\phi,\epsilon}+O(\lambda^{2})
  \right]
\label{eq-G-expansion-replica}
\eeqn
Note that $F_{m}[\hat{\rho},q]$ and $G_{m}[\phi,\epsilon]$ are related to each other by the Legendre transform. The functional integrations in \eq{eq-legendre-replica} can be done by the saddle point method for $N \gg 1$ which yield,
\beqn
-\beta F_{m}[\hat{\rho},q_{ab}] =
\int dz\phi^{*}(z)\hat\rho(z)
-\frac{d}{2}\sum_{a,b=1}^{m-1}\int dz \epsilon^{*}_{ab}(z)q_{ab}(z)\hat\rho(z)
-\beta G_{m}[\hat{\phi},\epsilon]
\eeqn
with the saddle point $(\phi^{*}(z),\epsilon^{*}(z))$ is determined by,
\beqn
\hat{\rho}(z)&=&\left. \frac{\delta (-\beta G_{m}[\phi,\epsilon_{ab}])}{\delta(-\phi(z))} \right|_{\phi=\phi^{*},\epsilon_{ab}=\epsilon_{ab}^{*}} \\
q_{ab}(z)\hat\rho(z)&=& \left. 
\frac{1}{d} \frac{\delta (-\beta G_{m}[\phi,\epsilon_{ab}]) }{\delta \epsilon_{ab}(z)}
\right |_{\phi=\phi^{*},\epsilon_{ab}=\epsilon_{ab}^{*}}
\label{eq-SP-replica}
\eeqn

In \eq{eq-G-expansion-replica}, similarly to \eq{eq-G}, we introduced a parameter $\lambda$
to organize the Mayer expansion. There
we also introduced the replicated Mayer function,
\beq
f^{m}_{ij}=e^{-\beta \sum_{a=1}^{m}v(r_{ij}^{a})}-1 \qquad
r_{ij}=|({\bf x}_{i}+{\bf u}_{i})-({\bf x}_{j}+{\bf u}_{j})|
\eeq
and the non-interacting part of the free-energy
or the free-energy of ideal-gas made of the 'molecules',
\beqn
e^{-\beta  G_{0,m}[\phi,\epsilon_{ab}]}&=&
\frac{1}{N!} e^{-N\beta  g_{0,m}[\phi,\epsilon_{ab}]}
\label{eq-g0-a}
\eeqn
with
\beqn
e^{-\beta  g_{0,m}[\phi,\epsilon_{ab}]} &=&
\left(\frac{m}{\lambda^{m}_{\rm th}}\right)^{d}
\int d^{d-1}x
\int dz e^{-\phi(z)}
\prod_{\mu=1}^{d}
\left \{\int
\prod_{a=1}^{m}d(u^{\mu})^{a} \delta(\sum_{a=1}^{m} (u^{\mu})^{a})
e^{\frac{d}{2}\sum_{a,b=1}^{m-1}\epsilon_{ab}(z)
  (u^{\mu})^{a}  (u^{\mu})^{b}}
\right \}
\prod_{a=1}^{m}
e^{-\beta U(z+(u^{d})_{a})}
\nonumber \\
&=& 
m^{d}
  S \int dz a(z)
\label{eq-g0-b}  
  \eeqn
where we introduced the 'activity',
  \beq
  a(z)=\frac{e^{-\phi(z)-\beta m U(z)}}{\lambda_{\rm th}^{d}}
    \left[
    \frac{(2\pi/\lambda^{2}_{\rm th})^{m-1}}{{\rm det}(-d\hat{\epsilon}^{m,m}(z))}
    \right]^{d/2}
    \eeq

    In \eq{eq-G-expansion-replica} we also introduced,
\beqn
\langle \ldots \rangle_{\phi,\epsilon}=
\frac{
\prod_{i=1}^{N} \int d^{d-1}x_{i}dz_{i}
e^{-\phi(z_{i})-\beta m U(z_{i})}
\int
\prod_{a=1}^{m}d(u^{\mu})^{a} \delta(\sum_{a=1}^{m} (u^{\mu})^{a})
e^{\frac{d}{2}\sum_{a,b=1}^{m-1}\epsilon_{ab}(z_{i})
  (u^{\mu})^{a}  (u^{\mu})^{b}}
  \ldots}{
  \prod_{i=1}^{N} \int d^{d-1}x_{i}dz_{i}
e^{-\phi((z_{i}))-\beta m U((z_{i}))}
\int
\prod_{a=1}^{m}d(u^{\mu})^{a} \delta(\sum_{a=1}^{m} (u^{\mu})^{a})
e^{\frac{d}{2}\sum_{a,b=1}^{m-1}\epsilon_{ab}(z_{i})
  (u^{\mu})^{a}  (u^{\mu})^{b}}
}
\qquad
\eeqn
which is the thermal average took within the non-interacting system.

Let us note that we replaced $U(z+(u^{d})_{a})$ by $U(z)$ dropping the correction terms due to $u^{d}_{a}$.
As we will see below $|{\bf u}|^{2} \sim O(1/d))$ so that the correction terms
can be neglected in $d \to \infty$ limit.

Again we treat the effect of interactions perturbatively
assuming the parameter $\lambda$ defined in \eq{eq-G-expansion-replica}
is small. Similarly to \eq{eq-legendre} we consider series
expansions,
\beqn
\phi^{*} &=& \phi^{*}_{0}+\lambda \phi^{*}_{1} + \frac{\lambda^{2}}{2} \phi^{*}_{2} + 
\qquad \epsilon_{ab}^{*} = (\epsilon^{*}_{0})_{ab}+\lambda (\epsilon^{*}_{1})_{ab}
+ \frac{\lambda^{2}}{2} (\epsilon^{*}_{2})_{ab} + \qquad \nonumber \\
G_{m}  &=& G_{0,m}+\lambda G_{1,m} + \frac{\lambda^{2}}{2} G_{2,m} + \ldots \qquad
F_{m}  = F_{0,m}+\lambda F_{1,m} + \frac{\lambda^{2}}{2} F_{2,m} + \ldots
\label{eq-legendre-transform-expansion-replica}
\eeqn
in terms of $\lambda$. We find,
\beqn
-\beta F_{0,m}[\hat{\rho},q_{ab}] &=&
\int dz\phi_{0}^{*}(z)\hat\rho(z)
-\frac{d}{2}\sum_{a,b=1}^{m-1}\int dz (\epsilon_{0}^{*})_{ab}(z)q_{ab}(z)\hat\rho(z)
-\beta G_{0,m}[\hat{\phi}_{0},(\epsilon_{0}^{*})_{ab}] \label{eq-F0m}\\
-\beta F_{1,m}[\hat{\rho},q_{ab}]&=&-\beta G_{1,m}[\phi_{0}^{*},(\epsilon_{0}^{*})_{ab}] \label{eq-F1m}
\eeqn

To work out $\phi_{0}^{*}$ and $\epsilon^{*}_{ab}$ we use
\eq{eq-g0-a} and \eq{eq-g0-b} in \eq{eq-SP-replica} and find,
\beqn
\hat{\rho}(z)&=&N\frac{a^{*}(z)}{\int dz a^{*}(z)}
\qquad a^{*}(z)=
\frac{e^{-\phi_{0}^{*}(z)-\beta m U(z)}}{\lambda_{\rm th}^{d}}
    \left[
    \frac{(2\pi/\lambda_{\rm th}^{2})^{m-1}}{{\rm det}(-d(\hat{\epsilon}_{0}^{*})^{m,m}(z))}
    \right]^{d/2}
\\
q_{ab}(z)&=&-((\hat\epsilon_{0}^{*})^{m,m})^{-1}_{ab}
\eeqn
using this in \eq{eq-F0m} we find the ideal gas part (entropic) part of the free-energy,
\beqn
\frac{-\beta F_{0,m}[\rho,q_{ab}]}{S}
&=&\int dz \rho(z)
\left[1-\ln \left(\lambda_{\rm th}^{d}\rho(z)\right) + d \ln m
  +\frac{(m-1)d}{2}\ln \left(\frac{2\pi e}{d}
  \right) \right]  \nonumber \\
&+&\int dz \rho(z)
\left[-\beta m U(z)+\frac{d}{2}\ln \left(
\frac{{\rm det}\;\hat{q}^{m,m}(z)}{\lambda_{\rm th}^{2(m-1)}}\right)
\right]
\label{eq-F0-replica}
\eeqn
We also used \eq{eq-rho-and-hat-rho}  which reads $\rho(z)=\hat{\rho}(z)/S$.

Now let us consider the effect of interactions as we did in the previous section. From \eq{eq-G-expansion-replica} we find,
\beq
-\beta G_{1,m}[\phi]=\sum_{i < j} \langle f_{m}(\{r^{a}_{ij}\}) \rangle_{\phi,\epsilon} 
=\frac{N(N-1)}{2}
\frac{\prod_{i=1}^{2} \int d^{d-1}x_{i}dz_{i} a(z_{i})
f^{m}_{ij}  
}{\prod_{i=1}^{2} \int d^{d-1}x_{i}dz_{i} a(z_{i})}
\eeq
Using this in the last equation of 
\eq{eq-legendre-transform-expansion-replica}
we find,
\beqn
\frac{-\beta F_{1,m}[\rho,q_{ab}]}{S}&=&\frac{1}{2}\int dz_{1} dz_{2} \rho(z_{1})\rho(z_{2})
\frac{1}{S}\int d^{d-1}x_{1}\int d^{d-1}x_{2}\langle f^{m}_{ij}
\rangle_{\epsilon^{*}_{0}(z_{1}),\epsilon^{*}_{0}(z_{2})} \nonumber \\
&=&  \frac{1}{2}(d-1)\Omega_{d-1}\int dz_{1} dz_{2} \rho(z_{1})\rho(z_{2})
\int dr_{12,\perp}r_{12,\perp}^{d-1}
\langle \langle f^{m}_{ij}
\rangle_{\epsilon^{*}_{0}(z_{1}),\epsilon^{*}_{0}(z_{2})}  \rangle_{\Omega}
\label{eq-F1-replica}
\eeqn
where we defined,
\beq
\langle \ldots \rangle_{\epsilon}
=\frac{
  \int
\prod_{a=1}^{m}d(u^{\mu})^{a} \delta(\sum_{a=1}^{m} (u^{\mu})^{a})
e^{\frac{d}{2}\sum_{a,b=1}^{m-1}\epsilon_{ab}(z_{i})
  (u^{\mu})^{a}  (u^{\mu})^{b}} \ldots
}{
  \int
\prod_{a=1}^{m}d(u^{\mu})^{a} \delta(\sum_{a=1}^{m} (u^{\mu})^{a})
e^{\frac{d}{2}\sum_{a,b=1}^{m-1}\epsilon_{ab}
  (u^{\mu})^{a}  (u^{\mu})^{b}}
}
\eeq
which is the thermal average of fluctuations within a molecule
and  the average over the solid angle $\Omega$ associated with the displacement vector ${\bf r}_{12,\perp}$,
\beq
\langle \ldots \rangle_{\Omega}=\frac{\int d\Omega \ldots }{
  \Omega
} \qquad \Omega=(d-1)\Omega_{d-1}
\label{eq-averaging-over-solid-angle}
\eeq

Now
\beqn
r^{2}_{12}&=&|({\bf x}_{1}+{\bf u}_{1})-({\bf x}_{2}+{\bf u}_{2})|^{2} \nonumber \\
 &=& (({\bf x}_{1})^{a}-({\bf x}_{2})^{a})^{2}
+(({\bf u}_{1})^{a}-({\bf u}_{2})^{a})^{2}
+2(({\bf u}_{1})^{a}-({\bf u}_{2})^{a})
\cdot (({\bf x}_{1})^{a}-({\bf x}_{2})^{a}) \nonumber \\
&=& r^{2}_{12,\perp}+
\underbrace{(z_{1}-z_{2})^{2}+(({\bf u}_{1})^{a}-({\bf u}_{2})^{a})^{2}
+2(({\bf u}_{1,\perp})^{a}-({\bf u}_{2,\perp})^{a})
\cdot {\bf r}_{12,\perp}}_{X^{a}}
\eeqn

Here we drooped $((u^{d}_{1})^{a}-(u^{d}_{2})^{a})(z_{1}-z_{2})$ 
anticipating $|{\bf u}|^{2} \sim O(1/d)$.

Now writing
\beq
f^{m}_{ij}=f^{m}(\{r_{12,\perp}^{2}+X^{a}\})=
\left. e^{\sum_{a} X^{a}\frac{\partial}{\partial y_{a}}}
f^{m}(\{r_{12,\perp}^{2}+y^{a}\}) \right |_{y^{a}=0}
\eeq
we find
\beqn
&&\langle \langle f^{m}_{ij}
\rangle_{\epsilon(z_{1}),\epsilon(z_{2})}  \rangle_{\Omega} \nonumber \\
&&= \langle \langle 
\left. e^{\sum_{a}X^{a}\frac{\partial}{\partial y_{a}}}
\rangle_{\epsilon(z_{1}),\epsilon(z_{2})}  \rangle_{\Omega}
f^{m}(r_{12,\perp}^{2}+y^{a}) \right |_{y^{a}=0} \nonumber \\
&& = e^{(z_{1}-z_{2})^{2} \sum_{a}\frac{\partial}{\partial y_{a}}}
\left \langle
e^{\sum_{a}(({\bf u}_{1})^{a}-({\bf u}_{2})^{a})^{2}\frac{\partial}{\partial y_{a}}}
\left \langle 
\left. e^{\sum_{a}[
2(({\bf u}_{1,\perp})^{a}-({\bf u}_{2,\perp})^{a})
\cdot {\bf r}_{12,\perp} ]\frac{\partial}{\partial y_{a}}}
 \right \rangle_{\Omega} \right \rangle_{\epsilon(z_{1}),\epsilon(z_{2})} 
 f^{m}(r_{12,\perp}^{2}+y^{a}) \right |_{y^{a}=0}   
 \label{eq-fij-replica-average}
 \eeqn

Let us pause for a moment to investigate
the averaging over the solid angle $\Omega$
of the unit vector $\hat{r}_{12,\perp}={\bf r}_{12,\perp}/r_{12,\perp}={\bf y}/\sqrt{d-1}$
defined in \eq{eq-averaging-over-solid-angle}.
We can notice that the average over the solid angle $\Omega$ can be done assuming
the vector $\hat{r}_{12,\perp}$ obey a Gaussian distribution in $d \to \infty$ limit,
\beq
P(\hat{r}_{12,\perp})=
\prod_{\mu=1}^{d}
\frac{
  e^{-\frac{(d-1)(\hat{r}^{\mu}_{12,\perp})^{2}}{2}}
}{\sqrt{2\pi/d}}
\eeq
This can be seen by writing
\beq
\langle \ldots \rangle_{\Omega}
=\Omega^{-1}\int \prod_{\mu=1}^{d-1}dx_{\mu} \delta(\sum_{\mu=1}^{d-1}x_{\mu}^{2}-1)
\propto \int \prod_{\mu=1}^{d-1}dy_{\mu} \delta(\sum_{\mu=1}^{d-1}y_{\mu}^{2}-(d-1))
=\int\frac{d\kappa}{2\pi}e^{-i\kappa (d-1)}\prod_{\mu=1}^{d-1}\int dy_{\mu}
e^{i\kappa y_{\mu}^{2}}\ldots
\nonumber
\eeq
In $d-1 \to \infty$ limit the integration over $\kappa$ can be done (formally)
by the saddle point method  so that different $y_{\mu}$'s can actually
be regarded as independent Gaussian random variables with zero mean and unit variance.
Based on this observation  we find,
\beqn
&& \ln \left \langle
e^{\sum_{a}(({\bf u}_{1})^{a}-({\bf u}_{2})^{a})^{2}
  \frac{\partial}{\partial y_{a}}}
\left \langle
 e^{\sum_{a}[
2(({\bf u}_{1,\perp})^{a}-({\bf u}_{2,\perp})^{a})
\cdot {\bf r}_{12,\perp} ]
      \frac{\partial}{\partial y_{a}}}
\right \rangle_{\Omega} \right\rangle_{\epsilon(z_{1}),\epsilon(z_{2})}  \nonumber \\
&=&
\ln \left \langle 
\exp \left [
  \sum_{a}
  (({\bf u}_{1})^{a}-({\bf u}_{2})^{a})^{2}\frac{\partial}{\partial y_{a}}  
  +
  \frac{(2r_{12,\perp})^{2}}{2(d-1)}
  \sum_{a,b}
  (({\bf u}_{1,\perp})^{a}-({\bf u}_{2,\perp})^{a})\cdot(({\bf u}_{1,\perp})^{b}-({\bf u}_{2,\perp})^{b})
   \frac{\partial}{\partial y_{a}}\frac{\partial}{\partial y_{b}}
\right]
\right\rangle_{\epsilon(z_{1}),\epsilon(z_{2})}  \nonumber \\
&=&
\frac{1}{2}
\sum_{a} (\alpha_{aa}(\hat{z}_{1})+\alpha_{aa}(\hat{z}_{2}))
\frac{2D^{2}}{d}
    \frac{\partial}{\partial y_{a}}
  +
\frac{1}{2}  \sum_{a,b}
     (
       \alpha_{ab}(\hat{z}_{1})+\alpha_{ab}(\hat{z}_{2})
       )
  \frac{2(r_{12,\perp})^{2}}{d-1}\frac{2D^{2}}{d}
       \frac{\partial}{\partial y_{a}}\frac{\partial}{\partial y_{b}}
\eeqn
In the  last equation we introduced $\alpha_{ab}$ such that
\beq
q_{ab}=\langle {\bf u}_{a} \cdot {\bf u}_{b}\rangle_{\epsilon} = \frac{D^{2}}{d}\alpha_{ab}
\label{eq-def-alpha_ab}
\eeq
In order to have sensible results in $d\to \infty$ limit we consider $\alpha_{ab} \sim O(1)$
which means $\langle u^{2} \rangle  \sim O(1/d)$.
With this scaling we could drop higher order
terms that appear in the cumulant expansion of $ \ln \langle \ldots \rangle_{\epsilon}$. Note also that $\langle {\bf u_\perp}_{a} \cdot {\bf u_\perp}_{b}\rangle_{\epsilon}=q_{ab}$ dropping $1/d^{2}$ correction.

Using this back in \eq{eq-fij-replica-average}
we find,
\beqn
\langle \langle f^{m}_{ij}
\rangle_{\epsilon(z_{1}),\epsilon(z_{2})}  \rangle_{\Omega} 
& =&
e^{
\frac{1}{2}
\sum_{a} (\alpha_{aa}(\hat{z}_{1})+\alpha_{aa}(\hat{z}_{2}))
    \frac{\partial}{\partial \xi_{a}}
  +
\frac{1}{2}  \sum_{a,b}
     (
       \alpha_{ab}(\hat{z}_{1})+\alpha_{ab}(\hat{z}_{2})
       )
              \frac{\partial^{2}}{\partial \xi_{a} \partial \xi_{b}}
}
\left .
 f^{m}\left(
 \left \{
 D^{2}\left(1+
   \frac{\xi_{a}+\frac{(\hat{z}_{1}-\hat{z}_{2})^{2}}{2}}{d-1}\right)^{2}
\right \}
\right) \right |_{\xi^{a}=0}   \nonumber \\
& =&
e^{
-\frac{1}{2}  \sum_{a,b} 
     \Delta_{ab}(\hat{z}_{1},\hat{z}_{2})
       \frac{\partial^{2}}{\partial \xi_{a}\partial \xi_{b}}
+\alpha_{d}(\hat{z}_{1},\hat{z}_{2})
\sum_{a}     \frac{\partial}{\partial \xi_{a}}
\left(\sum_{a}     \frac{\partial}{\partial \xi_{a}}+1\right)
}
\left .
 f^{m}\left(
 \left \{
 D^{2}\left(1+
   \frac{\xi_{a}+\frac{(\hat{z}_{1}-\hat{z}_{2})^{2}}{2}}{d-1}\right)^{2}
\right \}
\right) \right |_{\xi^{a}=0}
 \eeqn
 Here we used \eq{eq-xi} $r_{12,\perp}=D(1+\xi/(d-1))$,
 $\frac{\partial}{\partial y_{a}}=(d-1)/(2D^{2}) \frac{\partial}{\partial \xi_{a}}$ and the identity $f(x+a)=e^{a\frac{d}{dx}}f(x)$ for a generic function $f(x)$. We introduced
 \beq
 \Delta_{ab}(\hat{z}_{1},\hat{z}_{2})=2\alpha_{d}(\hat{z}_{1},\hat{z}_{2})-2\alpha_{ab}(\hat{z}_{1},\hat{z}_{2})
 \qquad
 \alpha(\hat{z}_{1},\hat{z}_{2})=\frac{\alpha_{ab}(\hat{z}_{1})+\alpha_{ab}(\hat{z}_{2})}{2}
 \label{eq-def-Delta-alpha}
 \eeq
 assuming
 \beq
 \qquad \alpha_{aa}(\hat{z}_{1},\hat{z}_{2})=\alpha_{\rm d}(\hat{z}_{1},\hat{z}_{2})
 \eeq
for $a=1,2,\ldots,m$.
We also used \eq{eq-z-and-scaled-z} which reads $z=(D/\sqrt{d})\hat{z}$. 
Then using these expressions back in \eq{eq-F1-replica} and using
\beq
\Omega_{d-1}=\sqrt{\frac{d}{2\pi}}\Omega_{d}
\eeq
we find
 \beqn
&& \frac{-\beta F_{1,m}[\rho,q_{ab}]}{S D/\sqrt{d}}  \nonumber \\
&=&  \frac{d}{2}\frac{\Omega_{d}}{d}D^{d} \int d\hat{z}_{1} \int d\hat{z}_{2}\rho(\hat{z}_{1})\rho(\hat{z}_{2})
\frac{e^{-\frac{(\hat{z}_{1}-\hat{z}_{2})^{2}}{2}}}{\sqrt{2\pi}}
\int d\xi e^{\xi}
e^{
-\frac{1}{2}  \sum_{a,b} 
     \Delta_{ab}(\hat{z}_{1},\hat{z}_{2})
       \frac{\partial^{2}}{\partial \xi_{a}\partial \xi_{b}}
}
 \left .
 f^{m}\left(
 \left \{
 D^{2}\left (1+ \frac{\xi_{a}}{d-1} \right)
\right \}
\right) \right |_{\xi^{a}=0}
\eeqn
In the last equation we performed integrations over parts with respect to $\xi$ to eliminate the term $\alpha_{d}...$ in \eq{eq-F1-replica}. We also used
  \eq{eq-rho-and-hat-rho}  which reads $\rho(\hat{z})=\hat{\rho}(z)/S$.

On the other hand the entropic part of the free-energy \eq{eq-F0-replica} can be rewritten as,
 \beqn
 \frac{-\beta F_{0,m}[\rho,q_{ab}]}{S D/\sqrt{d}} 
& = & \int d\hat{z} \rho(\hat{z})
\left[1-\ln \left(\lambda_{\rm th}^{d}\rho(\hat{z})\right) + d \ln m
  +\frac{(m-1)d}{2}\ln \left(\frac{2\pi e(D/\lambda_{\rm th})^{2}}{d^{2}}
  \right) \right]  \nonumber \\
&+&\int d\hat{z} \rho(\hat{z})
\left[-\beta m U(\hat{z})+\frac{d}{2}\ln \left(
{\rm det}\;\hat{\alpha}^{m,m}(\hat{z})\right)
\right]
\eeqn
Again we used \eq{eq-z-and-scaled-z} which reads $z=(D/\sqrt{d})\hat{z}$. 

To wrap up the results we find
\beqn
-\beta F &=&\frac{1}{m} \log  \int {\cal D}\rho(\hat{z})\int {\cal D}[q_{ab}(\hat{z})]
e^{-\beta F_{m}[\rho(\hat{z}),q_{ab}(\hat{z})]} \nonumber \\
&=& \frac{1}{m} (-\beta F_{m}[\rho^{*}(\hat{z}),q^{*}_{ab}(\hat{z})])
\eeqn
with
\beqn
& \frac{-\beta F_{m}[\rho,q_{ab}]}{SD/\sqrt{d}}=
\int d\hat{z} \rho(\hat{z})
\left[1-\ln \left(\lambda_{\rm th}^{d}\rho(\hat{z})\right) + d \ln m
  +\frac{(m-1)d}{2}\ln \left(\frac{2\pi e(D/\lambda_{\rm th})^{2}}{d^{2}}
  \right) \right]  \nonumber \\
&+\int d\hat{z} \rho(\hat{z})
\left[-\beta m U(\hat{z})+\frac{d}{2}\ln \left(
{\rm det}\;\hat{\alpha}^{m,m}(\hat{z})\right)
\right] \nonumber \\
&+ \frac{d}{2}\frac{\Omega_{d}}{d}D^{d} \int d\hat{z}_{1} \int d\hat{z}_{2}\rho(\hat{z}_{1})\rho(\hat{z}_{2})
\frac{e^{-\frac{(\hat{z}_{1}-\hat{z}_{2})^{2}}{2}}}{\sqrt{2\pi}}
\int d\xi e^{\xi}
e^{
-\frac{1}{2}  \sum_{a,b} 
     \Delta_{ab}(\hat{z}_{1},\hat{z}_{2})
       \frac{\partial^{2}}{\partial \xi_{a}\partial \xi_{b}}
}
 \left .
 f^{m}\left(
 \left \{
 D^{2}\left (1+ \frac{\xi_{a}}{d-1} \right)
\right \}
\right) \right |_{\xi^{a}=0}
\;\;\;
\label{eq-F-large-d-glass}
\eeqn
where $\rho^{*}(\hat{z})$ and $q_{ab}^{*}(\hat{z})$ are solutions of the saddle point equations,
\beqn
0&=&\frac{\delta}{\delta \rho(\hat{z})}
\left [ (-\beta F_{m}[\rho,q_{ab}]) +\beta \mu \left(
  \int d\hat{z} \rho(\hat{z})-\frac{N}{S}\frac{\sqrt{d}}{D}\right) \right] \nonumber \\
0&=& \frac{\delta}{\delta q_{ab}(\hat{z})}
(-\beta F_{m}[\rho,q_{ab}])
\eeqn
The 1st equation yields the chemical potential,
\beqn
-\beta m \mu  &=&
\left. \frac{\delta}{\delta \rho(\hat{z}_{1})}
\frac{-\beta m F[\rho]}{
  SD/\sqrt{d}}
\right |_{\rho=\rho^{*}} \nonumber \\
&=&-\ln (\lambda_{\rm th}^{d}\rho^{*}(\hat{z}_{1}))
+ (-\beta m U(\hat{z}_{1})) +\frac{d}{2} \ln \left(
{\rm det}\;\hat{\alpha}^{m,m}(\hat{z})\right)
\nonumber \\
&& +d \frac{\Omega_{d}}{d} D^{d}
\int d\hat{z}_{2} \rho^{*}(\hat{z}_{2})
\frac{e^{-\frac{(\hat{z}_{1}-\hat{z}_{2})^{2}}{2}}}{\sqrt{2\pi}}
\int_{-\infty}^{\infty}d\xi e^{\xi}
e^{
-\frac{1}{2}  \sum_{a,b} 
     \Delta_{ab}(\hat{z}_{1},\hat{z}_{2})
       \frac{\partial^{2}}{\partial \xi_{a}\partial \xi_{b}}
}
f^{m}\left[D^{2}
  \left.
  \left(
1+\frac{\xi_{a}}{d}
  \right)^{2}
  \right]
\right |_{\xi^{a}=0}
\label{eq-mu-replica}
\eeqn

Using the scaled volume fraction $\hat\varphi$ defined in \eq{eq-hat-var-phi} we can write
\beqn
& \frac{-\beta F_{m}[\hat\varphi,q_{ab}]}{SD/\sqrt{d}}\frac{\Omega_{d}D^{d}}{d}=
\int d\hat{z}\hat{\varphi}(\hat{z})
\left [1-
  \ln \left (\hat{\varphi}(\hat{z})(d/\Omega_{d})(\lambda_{\rm{th}}/D)^{d} \right)+ (-\beta m \hat{U}_{0}(\hat{z})) \right]
\nonumber \\
& +d \left\{ \int d\hat{z}\hat{\varphi}(\hat{z})
\left[
(-\beta m \hat{U}_{1}(\hat{z}))+
  \frac{1}{2}
\ln \left(
    {\rm det}\;\hat{\alpha}^{m,m}(\hat{z})\right)\right] \right. \nonumber \\
& \left.    +\frac{1}{2}
    \int d\hat{z}_{1} \int d\hat{z}_{2}\hat\varphi(\hat{z}_{1})\hat\varphi(\hat{z}_{2})
\frac{e^{-\frac{(\hat{z}_{1}-\hat{z}_{2})^{2}}{2}}}{\sqrt{2\pi}}
\int d\xi e^{\xi}
e^{
-\frac{1}{2}  \sum_{a,b} 
     \Delta_{ab}(\hat{z}_{1},\hat{z}_{2})
       \frac{\partial^{2}}{\partial \xi_{a}\partial \xi_{b}}
}
 \left .
 f^{m}\left(
 \left \{
 D^{2}\left (1+ \frac{\xi_{a}}{d-1} \right)
\right \}
\right) \right |_{\xi^{a}=0}
\;\;\;
\right\}
\label{eq-F-large-d-glass-by-varphi}
\eeqn
Similarly the chemical potential can be expressed as,
\beqn
-\beta m \mu
&=&-\ln[(\lambda_{\rm th}/D)^{d}(d/\Omega_{d})]
-\ln \hat\varphi(\hat{z}_{1})
+  (-\beta m \hat{U}_{0}(\hat{z}_{1})) \nonumber \\
&+& d\left\{
 (-\beta m \hat{U}_{1}(\hat{z}_{1}))
+\frac{1}{2} \ln \left(
{\rm det}\;\hat{\alpha}^{m,m}(\hat{z})\right) \nonumber \right. \\
&& \left. 
+\int d\hat{z}_{2} \hat{\varphi}(\hat{z}_{2})
\frac{e^{-\frac{(\hat{z}_{1}-\hat{z}_{2})^{2}}{2}}}{\sqrt{2\pi}}
\int_{-\infty}^{\infty}d\xi e^{\xi}
e^{
-\frac{1}{2}  \sum_{a,b} 
     \Delta_{ab}(\hat{z}_{1},\hat{z}_{2})
       \frac{\partial^{2}}{\partial \xi_{a}\partial \xi_{b}}
}
\left . 
f^{m}\left[D^{2}\left(
1+\frac{\xi_{a}}{d}
  \right)^{2}
  \right] \right|_{\xi_{a}=0} \right \} \qquad\qquad
\label{eq-chemical-potential-replica}
\eeqn

It is useful to recall the special case of uniform density profile $\hat{\varphi}(\hat{z})=\hat{\varphi}$
and spatially uniform glass order parameter $\Delta_{ab}(\hat{z})=\Delta_{ab}$. 
In this case, the free-energy becomes,
\beqn
& \frac{-\beta F_{m}[\hat\varphi,q_{ab}]}{N}
    =   1-\ln{\hat\varphi}-\ln {[ (\lambda_{\rm{th}}/D)^{d}(d/\Omega_{d})]}
      + d\ln{m}+\frac{(m-1)d}{2}\ln{\frac{2\pi e D^2}{d^2\lambda_{\rm{th}}^{2}}} 
\nnum\\
&    +\frac{d}{2}
\left \{
 \ln \left(
{\rm det}\;\hat{\alpha}^{m,m}(\hat{z})\right)
    +\hat\varphi
    \int_{-\infty}^{\infty} d\xi 
e^{
-\frac{1}{2}  \sum_{a,b} 
     \Delta_{ab}
       \frac{\partial^{2}}{\partial \xi_{a}\partial \xi_{b}}
}
\left . 
f^{m}\left[D^{2}\left(
1+\frac{\xi_{a}}{d}
  \right)^{2}
  \right] \right|_{\xi_{a}=0}
\right \}
\label{eq-free-ene-uniform-general}
\eeqn
Here we have switched off the external potential $U$.

\subsection{Compression}
\label{sec-compression-replica}

Now let us extend the analysis in  sec.~\ref{sec-compression-liquid}
to discuss compression (or decompression) of our system in the glassy states.
In the replicated system we may consider compressing each replica differently using
$\eta_{a}$ ($a=1,2,\ldots,m$).

Then the expression of the free-energy \eq{eq-F-total-glass} becomes
\beqn
-\beta F(\{\eta_{a}\}) &=& \frac{1}{m}\ln 
\frac{1}{(N!)^{m}} \prod_{a \in {\cal C}}
\int_{\{V(\eta_{a})\}} \prod_{i=1}^{N}\frac{d^{d-1}(x^{a}_{i})'}{\lambda_{\rm th}^{d-1}}\frac{d(z^{a}_{i})'}{\lambda_{\rm th}}e^{-\beta \sum_{i< j}v((r_{ij}^{a})')-\beta \sum_{i} U((z_{i}^{a})')}
\nonumber \\
&=& \frac{1}{m}\ln 
\frac{1}{(N!)^{m}}\prod_{a \in {\cal C}}
\underbrace{\left(1-\frac{\eta_{a}}{d}\right)^{d}}_{\mbox{$e^{-\eta_{a}}$ in $d\to\infty$}}
\int_{\{V(0)\}} \prod_{i=1}^{N} \frac{d^{d-1}x^{a}_{i}}{\lambda_{\rm th}^{d-1}}\frac{dz^{a}_{i}}{\lambda_{\rm th}}
e^{-\beta \sum_{i<j} v(r_{ij}(1-\eta_{a}/d))-\beta \sum_{i=1}^{N}U(z_{i}(1-\eta_{a}/d))}
\qquad \qquad
\eeqn
Then the free-energy functional \eq{eq-F-large-d-glass} becomes
\beqn
& -\frac{\beta F_{m}[\rho,q_{ab},\{\eta_{a}]\}}{
  SD/\sqrt{d}} =
\int d\hat{z} \rho(\hat{z})
\left[1-\sum_{a=1}^{m}\eta_{a}-\ln \left(\lambda_{\rm th}^{d}\rho(\hat{z})\right) + d \ln m
  +\frac{(m-1)d}{2}\ln \left(\frac{2\pi e(D/\lambda_{\rm th})^{2}}{d^{2}}
  \right) \right]  \nonumber \\
&+\int d\hat{z} \rho(\hat{z})
\left[-\beta m U(\hat{z}(1-\eta_{a}/d))+\frac{d}{2}\ln \left(
{\rm det}\;\hat{\alpha}^{m,m}(\hat{z})\right)
\right] \nonumber \\
&+ \frac{d}{2}\frac{\Omega_{d}}{d}D^{d} \int d\hat{z}_{1} \int d\hat{z}_{2}\rho(\hat{z}_{1})\rho(\hat{z}_{2})
\frac{e^{-\frac{(\hat{z}_{1}-\hat{z}_{2})^{2}}{2}}}{\sqrt{2\pi}}
\int d\xi e^{\xi}
e^{
-\frac{1}{2}  \sum_{a,b} 
     \Delta_{ab}(\hat{z}_{1},\hat{z}_{2})
       \frac{\partial^{2}}{\partial \xi_{a}\partial \xi_{b}}
}
 \left .
 f^{m}\left(
 \left \{
 D^{2}\left (1+ \frac{\xi_{a}}{d-1}-\frac{\eta_{a}}{d} \right)
\right \}
\right) \right |_{\xi^{a}=0}
\;\;\;
\label{eq-F-large-d-eta-glass}
\eeqn

We can compute the pressure assuming uniform deformation $\eta_{a}=\eta$ $(a=1,2,\ldots,m)$
and using \eq{eq-pressure}.
Here we omitted the contribution from the external potential.
The reduced pressure is obtained as,
\beqn
&&p=
-\frac{1}{m}\frac{\partial}{\partial \eta} \frac{-\beta F_{m}[\rho,q_{ab},\eta]}{SD/\sqrt{d}\int d\hat{z}\rho(\hat(z))} =1+\nonumber \\
&& 
\frac{1}{m}\left[\int d\hat{z} \hat{\rho}(\hat{z})\right]^{-1}
\left[
  \frac{d}{2} \frac{\Omega_{d}}{d} D^{d}
\int d\hat{z}_{1} \rho(\hat{z}_{1})
\int d\hat{z}_{2} \rho(\hat{z}_{2})
\frac{e^{-\frac{(\hat{z}_{1}-\hat{z}_{2})^{2}}{2}}}{\sqrt{2\pi}}
  \left (-\frac{\partial}{\partial \eta} \right)
\int_{-\infty}^{\infty}d\xi e^{\xi} e^{
-\frac{1}{2}  \sum_{a,b} 
     \Delta_{ab}(\hat{z}_{1},\hat{z}_{2})
       \frac{\partial^{2}}{\partial \xi_{a}\partial \xi_{b}}
}
f\left[D^{2}\left(
1+\frac{\xi}{d}-\frac{\eta}{d}
\right)^{2}\right]
\right] \qquad \qquad
\label{eq-reduced-pressure-replica}
\eeqn
with $\rho=N/V$.
We have $N=(SD/\sqrt{d})\int d\hat{z} \rho(\hat{z})$
(see \eq{eq-normalization-rho-mod}).
It can be seen that it verifies
\beq
p\int d\hat{z}\rho(\hat{z})= \int d\hat{z} \rho(\hat{z}) \beta \mu -\frac{\beta F}{SD/\sqrt{d}}
\eeq
which is equivalent to the thermodynamic relation $PV=\mu N-F$.

\section{One step RSB solution}
\label{appendix-1RSB}

\subsection{1RSB ansatz}
\label{appendix-1RSB-ansatz}

As we described at the beginning of sec~\ref{sec-basic-setup-replicated-liquid-theory} we are considering
'molecular liquid' made of $m$ replicas.
For the space dependent order parameter $\alpha_{ab}(\hat{z})$
with $a=1,2,\ldots,m$ and $b=1,2,\ldots,m$, we consider the simplest ansatz,
\beq
\alpha_{ab}(\hat{z})=(\alpha_{d}(\hat{z})+\alpha(\hat{z}))\delta_{ab}-\alpha(\hat{z})
\eeq
This ansatz reflects the symmetry of the system under permutations of the replicas $1,2,\ldots,m$.
This is the so called replica symmetry so that this ansatz
may be called as a replica symmetric (RS) ansatz \cite{1130848328245474560}. In the present paper we prefer to call this
ansatz as a one-step replica symmetry broken (1RSB) ansatz because we are considering a realization of a molecular
liquid state where the replica symmetry involving all replicas $1,2,...,n$ is reduced down to that
within a molecule $1,2,\ldots,m$ as we described in sec~\ref{sec-basic-setup-replicated-liquid-theory}.

Because of the sum rule \eq{eq-sum-rule-q} and \eq{eq-def-alpha_ab} we find
$\alpha_{d}(\hat{z})$ becomes $\alpha_{d}(\hat{z})=(m-1)\alpha(\hat{z})$.
The we can rewrite the ansatz as,
\beq
\alpha_{ab}(\hat{z})=(mI_{ab}-1)\alpha(\hat{z})
\label{eq-1RSB-ansatz}
\eeq
Equivalently using $\Delta_{ab}(\hat{z})$ defined in \eq{eq-def-Delta-alpha} the ansatz can be
written also as,
\begin{align}
  \Delta_{ab}(\hat{z})=\Delta(\hat{z})(1-I_{ab}).
\label{eq-1RSB-ansatz-Delta}  
\end{align}
with $\Delta(\hat{z})=2(\alpha_d(\hat{z})+\alpha(\hat{z}))=2m\alpha(\hat{z})$.

\subsection{Free energy}

Now let us evaluate the free-energy \eq{eq-F-large-d-glass} using the 1RSB ansatz.
In the entropic part of the free energy we find,
\begin{align}
    \ln \det \hat{\alpha}^{m,m}(\hat{z})
    =(m-1)\ln{(m\alpha(\hat{z}))}-\ln{m}
    =(m-1)\ln{\frac{\Delta(\hat{z})}{2}}-\ln{m}.
\end{align}

In the interaction part of the free-energy we find,
\begin{align}
    -\mathcal{F}_{\rm{int}}(\Delta(z,z'))
    &=\int^{\infty}_{-\infty}d\xi e^{\xi}
    e^{-\frac{1}{2}\sum_{ab}\Delta(\hat{z},\hat{z}')\partial_{\xi_a}\partial_{\xi_b}}\left. \qty[\prod_{a}e^{-\beta v\qty(D^2\qty(1+\frac{\xi_a}{d})^2)}-1]
    \right|_{\qty{\xi_a=\xi}}\nnum\\
    &=\int^{\infty}_{-\infty}d\xi e^{\xi}
    \qty[e^{-\frac{1}{2}\Delta(\hat{z},\hat{z}') \partial_{\xi}^2}
    \qty(e^{\frac{1}{2}\Delta(\hat{z},\hat{z}')}e^{-\beta v(D^2(1+\frac{\xi}{d})^2)})^m-1]\nnum\\
    &=\int^{\infty}_{-\infty}d\xi e^{\xi}
    \qty[e^{-\frac{1}{2}\Delta(\hat{z},\hat{z}') \partial_{\xi}^2}
    g^m\qty(\xi,\Delta(\hat{z},\hat{z}'))-1]\nnum\\
    &=\int^{\infty}_{-\infty}d\xi
    e^{\xi-\frac{1}{2}\Delta(\hat{z},\hat{z}')}
    \qty[g^m\qty(\xi,\Delta(\hat{z},\hat{z}'))-1]
\end{align}
where
\beq
\Delta(\hat{z},\hat{z}')=\frac{\Delta(\hat{z})+\Delta(\hat{z}')}{2}.
\eeq
We have also introduced
\beq
g\qty(\xi,\Delta) = e^{\frac{1}{2}\Delta\partial_{\xi}^{2}}e^{-\beta v(D^2(1+\xi/d)^2)}
=\int \mathcal{D}w e^{-\beta v\qty(D^2\qty(1+\frac{\xi+\sqrt{\Delta}w}{d})^2)}
    \label{eq-g-1RSB}
    \eeq
    \blue{
To derive the above equations we used the ideitities
   \eq{eq-diff-formulae},
    \eq{eq-diff-gaussian-integral-conversion}
      and the short-hand notation \eq{eq-def-Dz}}.

To sum up, we obtain the free energy within the 1RSB ansatz as,
\beqn
    \label{eq_free_energy_1RSB_simply_labeld}
    &\frac{-\beta F_{m}^{\rm{1RSB}} \qty[\qty{\Delta(\hat{z})}]}{S}\frac{\sqrt{d}}{D}
    =
    \int d\hat{z} \rho(\hat{z})\qty{
      1-\ln{(\rho(\hat{z}) \lambda_{\rm{th}}^{d})} + d\ln{m}+\frac{(m-1)d}{2}\ln{\frac{2\pi e D^2}{d^2\lambda_{\rm{th}}^{2}}}
   +(-\beta m \hat{U}_{0}(\hat{z}))      
    }
\nnum\\
    +&\frac{d}{2}
    \qty{
      \int d\hat{z} \rho(\hat{z})
    \qty[2(-\beta m \hat{U}_{1}(\hat{z}))+     (m-1)\ln{\frac{\Delta(\hat{z})}{2}}-\ln{m}]
    +\frac{\Omega_{d}D^d}{d}
    \int d\hat{z} \rho(\hat{z})
    \int d\hat{z}' \rho(\hat{z}')
\frac{e^{-\frac{(\hat{z}_{1}-\hat{z}_{2})^{2}}{2}}}{\sqrt{2\pi}}
\qty(-\mathcal{F}_{\rm{int}}\qty(\Delta(\hat{z},\hat{z}')))}. \qquad
    \label{eq-free-ene-1RSB-version1}
    \eeqn
    Or equivalently
\beqn
    \label{eq_free_energy_1RSB_simply_labeld_vaphi}
    &\frac{-\beta F_{m}^{\rm{1RSB}} \qty[\qty{\Delta(\hat{z})}]}{S}\frac{\sqrt{d}}{D}\frac{\Omega_{d}D^{d}}{d}
    =   \int d\hat{z} \hat\varphi(\hat{z})
    \qty{1-\ln{(\hat\varphi(\hat{z}) (d/\Omega_{d})(\lambda_{\rm{th}}/D)^{d})}
      + d\ln{m}+\frac{(m-1)d}{2}\ln{\frac{2\pi e D^2}{d^2\lambda_{\rm{th}}^{2}}}
      +(-\beta m \hat{U}_{0}(\hat{z}))
    },
\nnum\\
    +&\frac{d}{2}
    \qty{
     \int d\hat{z} \hat\varphi({z})
    \qty[2(-\beta m \hat{U}_{1}(\hat{z}))+(m-1)\ln{\frac{\Delta(\hat{z})}{2}}-\ln{m}]
    +
    \int d\hat{z}_{1} \hat\varphi(\hat{z}_{1})
    \int d\hat{z}_{2} \hat\varphi(\hat{z}_{2})
    \frac{e^{-\frac{(\hat{z}_{1}-\hat{z}_{2})^{2}}{2}}}{\sqrt{2\pi}}
    \qty(-\mathcal{F}_{\rm{int}}\qty(\Delta(\hat{z}_{1},\hat{z}_{2})))} \nonumber \\
    &={\rm const}+\frac{d}{2}(m-1)(-\beta V)(\{ \Delta(\hat{z})\})+O((m-1)^{2})
    \label{eq-free-ene-1RSB}
    \eeqn
    In the last equation the term 'const' means contributions independent of $\Delta(\hat{z})$ and we introduced,
    \beq
    -\beta V(\{ \Delta(\hat{z})\})=\int d \hat{z}\hat\varphi(\hat{z})
    \ln \Delta(\hat{z})-
    \int d\hat{z}_{1} \hat\varphi(\hat{z}_{1})
    \int d\hat{z}_{2} \hat\varphi(\hat{z}_{2})
    \frac{e^{-\frac{(\hat{z}_{1}-\hat{z}_{2})^{2}}{2}}}{\sqrt{2\pi}}
    \int_{-\infty}^{\infty} d\xi e^{\xi-\frac{1}{2}\Delta(\hat{z}_{1},\hat{z}_{2})}
    g(\xi,\Delta(\hat{z}_{1},\hat{z}_{2}))
    \ln g(\xi,\Delta(\hat{z}_{1},\hat{z}_{2}))
    \eeq
    which is the so called Franz-Parisi's potential.

\subsection{Equation of states}

The integration over $q_{ab}$ \eq{eq-partition-function-zm} can be done by the saddle point method.
The saddle point is found by solving,
\begin{align}
    0=\fdv{\Delta(\hat{z})}( -\beta F_m^{\rm 1RSB}[\{\Delta(\hat{z})\}])
\end{align}
which yields a self-consistent equation of $\Delta(\hat{z})$,
\begin{align}
   \label{append_self_consistent_eq_most_general}
    0
    &=\frac{1}{\Delta(\hat{z})}-\frac{m}{2}\int \frac{d\hat{z}'}{\sqrt{2\pi}}
    \hat{\varphi}(\hat{z}')
    e^{-\frac{(\hat{z}-\hat{z}')^2}{2}}
    \int_{-\infty}^{\infty} d\xi e^{\xi-\frac{1}{2}\Delta(\hat{z},\hat{z}')}g^{m-2}(\xi,\Delta(\hat{z},\hat{z}'))\qty(g'(\xi,\Delta(\hat{z},\hat{z}')))^{2}\nnum\\
    &=\frac{1}{\Delta(\hat{z})}-\frac{m}{2}\int \frac{d\hat{z}'}{\sqrt{2\pi}}
    \hat{\varphi}(\hat{z}')
    e^{-\frac{(\hat{z}-\hat{z}')^2}{2}}
    \int_{-\infty}^{\infty} d\xi e^{\xi-\frac{1}{2}\Delta(\hat{z},\hat{z}')}g^{m}(\xi,\Delta(\hat{z},\hat{z}'))\qty(f'(\xi,\Delta(\hat{z},\hat{z}')))^{2}
\end{align}
where we introduced
\beq
f(\xi,\Delta(\hat{z},\hat{z}'))=-\ln{g(\xi,\Delta(\hat{z},\hat{z}'))} 
\eeq

\subsection{Uniform system}
\label{sec-uniform-system}

Let us recall the special case of uniform density profile $\hat{\varphi}(\hat{z})=\hat{\varphi}$ and spatially uniform glass order parameter $\Delta(\hat{z})=\Delta$. In this case, the free-energy \eq{eq_free_energy_1RSB_simply_labeld_vaphi}
becomes, switching off the external potential $U$, 
\beqn
    &\frac{-\beta F_{m}^{\rm{1RSB}}(\Delta)}{N}
    =   1-\ln{\hat\varphi} -\ln {[(\lambda_{\rm{th}}/D)^{d}(d/\Omega_{d})]}
      + d\ln{m}+\frac{(m-1)d}{2}\ln{\frac{2\pi e D^2}{d^2\lambda_{\rm{th}}^{2}}} 
\nnum\\
&    +\frac{d}{2}
\left \{
    \qty[(m-1)\ln{\frac{\Delta}{2}}-\ln{m}]
    +\hat\varphi
\int_{-\infty}^{\infty} d\xi e^{\xi-\frac{1}{2}\Delta}\left[g^{m}(\xi,\Delta)-1\right]
\right \} \nonumber \\
&= {\rm const}+\frac{d}{2}
(m-1)(-\beta V)(\Delta)+ O ((m-1)^{2})
\eeqn
with $g(\xi,\Delta)$ defined in \eq{eq-g-1RSB}
and the term 'const' representing contributions independent of $\Delta$ and
\beq
-\beta V(\Delta)=
  \ln \Delta-\hat\varphi\int_{-\infty}^{\infty} d\xi e^{\xi-\frac{1}{2}\Delta} g(\xi,\Delta)\ln g(\xi,\Delta)
  \eeq
  is the Franz-Parisi's potential.
  
On the other hand  the equation of state
\eq{append_self_consistent_eq_most_general} becomes for the uniform glass state,
\beq
0=\blue{\frac{\partial (-\beta V(\Delta))}{\partial \Delta}=}\frac{1}{\Delta}-\frac{m}{2}\hat\varphi
\int_{-\infty}^{\infty} d\xi e^{\xi-\frac{1}{2}\Delta}g^{m}(\xi,\Delta)\qty(f'(\xi,\Delta))^{2}
\label{append_self_consistent_eq_most_uniform}
\eeq
\blue{Taking another derivative we find the Hessian of the uniform system}
\blue{
\beq
\frac{\partial^{2} (\beta V(\Delta))}{\partial \Delta^{2}}=
\frac{1}{\Delta^{2}}-\frac{m}{2}\hat\varphi X(\Delta)
\label{eq-hessian-bulk}
 \eeq
  with
  \beqn
X(\Delta)&=&-\frac{\partial}{\partial \Delta}\int d\xi e^{-\frac{\Delta}{2}\frac{\partial^{2}}{\partial \xi^{2}}}
g^{m}(\xi,\Delta)(f'(\xi,\Delta))^{2} \nonumber \\
&=& \frac{1}{2} \int d\xi e^{-\frac{\Delta}{2}\frac{\partial^{2}}{\partial \xi^{2}}}
\left[2(f^{''}(\xi))^{2}+(m-1)(-4f''(\xi))(f'(\xi))^{2}+m(m-1)(f'(\xi))^{4}\right]
\label{eq-def-X}
\eeqn
}

\subsection{Dynamical transition in uniform system}
\label{sec-dynamical-transition-in-uniform-system}

Solution to the equation of state \eq{append_self_consistent_eq_most_uniform}
has been studied in detail in previous works.
For instance in the the case of hard-spheres one finds non-trivial solutions
$\infty > \Delta > 0$ for high enough densities,
\beq
\Delta=\Delta_{\rm d}(m) - {\rm const} \sqrt{\hat\varphi-\hat\varphi_{\rm d}(m)}
\eeq
where $\hat\varphi_{\rm d}(m)$ is the so called dynamical transition density.
The dynamical transition point can be considered
as a sort of a spinodal point: 
the non-trivial solution associated with a local minimum of the
free-energy, which exists at higher densities, disappears there.

Close to the dynamical transition point we may expand the Franz-Parisi's potential as
\beq
-\beta V(\varphi,\Delta)=-\beta V_{0}(\varphi)+A(\hat\varphi)(\Delta-\Delta_{\rm d})+\frac{B(\hat\varphi)}{2!}(\Delta-\Delta_{\rm d})^{2}
+\frac{C(\hat\varphi)}{3!}(\Delta-\Delta_{\rm d})^{3}+\ldots
\eeq
with
\beqn
A(\hat\varphi)=A_{0}+A_{1}(\hat\varphi-\hat\varphi_{\rm d})+\ldots  \quad
B(\hat\varphi)=B_{0}+B_{1}(\hat\varphi-\hat\varphi_{\rm d})+\ldots  \quad
C(\hat\varphi)=C_{0}+C_{1}(\hat\varphi-\hat\varphi_{\rm d})+\ldots 
\eeqn
At the saddle point the 1st derivative must vanish (the equation of state
\eq{append_self_consistent_eq_most_uniform} ),
\beqn
0=\frac{\partial}{\partial \Delta}(-\beta V)(\hat\varphi,\Delta)=A(\hat\varphi)+B(\hat\varphi)(\Delta-\Delta_{\rm d})
+\frac{C(\hat\varphi)}{2}(\Delta-\Delta_{\rm d})^{2}+\ldots
\label{eq-eqs-around-dynamical-transition}
\eeqn
and the 2nd derivative (Hessian) is obtained as,
\beq
\frac{\partial^{2}}{\partial \Delta^{2}}(\beta V)(\hat\varphi,\Delta)=
-B(\hat\varphi)-C(\hat\varphi)(\Delta-\Delta_{\rm d})+\ldots
\label{eq-hessian-around-dynamical-transition}
\eeq

Considering \eq{eq-eqs-around-dynamical-transition} at $\hat\varphi=\hat\varphi_{\rm d}$
we find $A_{0}=0$. We also note that 2nd derivative must vanish at
$\hat\varphi=\hat\varphi_{\rm d}$
since it is a spinodal point as stated above, which implies $B_{0}=0$.
Using these observations in \eq{eq-eqs-around-dynamical-transition} we find
\beq
\Delta=\Delta_{\rm d}-\sqrt{-\frac{2A_{1}}{C_{0}}(\hat\varphi-\hat\varphi_{\rm d})}
-\frac{B_{1}}{C_{0}}(\hat\varphi-\hat\varphi_{\rm d})+
O((\hat\varphi-\hat\varphi_{\rm d})^{3/2})
\label{eq-order-parameter-scaling}
\eeq
Then this implies
\beq
\frac{\partial^{2}}{\partial \Delta^{2}}(\beta V)(\hat\varphi,\Delta)=
\sqrt{-2A_{1}C_{0}(\hat\varphi-\hat\varphi_{\rm d})}+
O((\hat\varphi-\hat\varphi_{\rm d})^{3/2})
\label{eq-scaling-hessian-bulk}
\eeq
at the saddle point close to the dynamical transition density $\hat\varphi_{\rm d}$.

%

\subsection{Longitudinal Hessian}
\label{sec-longtidudinal-mode-of-hessian}

In order to study the stability of the solutions to the saddle point equation we have to examine
the Hessian matrix. In the present paper we limit ourselves to what is called as 'longitudinal mode',
\beqn
-M(\hat{z}_{1},\hat{z}_{2})&=&\frac{\partial}{\partial \Delta(\hat{z}_{1})}\frac{\partial}{\partial \Delta(\hat{z}_{2})}
\frac{-\beta F_{m}^{\rm{1RSB}} \qty[\qty{\Delta(\hat{z})}]}{S D/\sqrt{d}} \frac{\Omega_d D^{d}}{d}\nonumber \\
&=& \frac{d}{2}(m-1)\hat{\varphi}(\hat{z}_{1})
\left[
  -\frac{1}{\Delta(\hat{z}_{1})^{2}}\delta(\hat{z}_{1}-\hat{z}_{2})
  +\frac{m}{2}\int d\hat{z} \hat{\varphi}(\hat{z})\frac{e^{-\frac{(\hat{z}_{1}-\hat{z})^{2}}{2}}}{\sqrt{2\pi}}
  \frac{\delta(\hat{z}_{1}-\hat{z}_{2})+\delta(\hat{z}-\hat{z}_{2})}{2}X(\Delta(\hat{z}_{1},\hat{z}))
  \right]\qquad
\eeqn
\blue{with $X(\Delta)$ defined in \eq{eq-def-X}.}

Then the free-energy around a glass state characterized with $\Delta^{*}(\hat{z})$
(which verifies the equation of state \eq{append_self_consistent_eq_most_general}) can be expanded as
\beq
\frac{-\beta F_{m}^{\rm{1RSB}} \qty[\qty{\Delta(\hat{z})}]}{S D/\sqrt{d}} \frac{\Omega_d D^{d}}{d}
=
\frac{-\beta F_{m}^{\rm{1RSB}} \qty[\qty{\Delta^{*}(\hat{z})}]}{S D/\sqrt{d}} \frac{\Omega_d D^{d}}{d}
-\frac{1}{2}\int d\hat{z}_{1}d\hat{z}_{2}M(\hat{z}_{1},\hat{z}_{2})\delta \Delta(\hat{z}_{1})\delta \Delta(\hat{z}_{2})+ \ldots
\label{eq-expansion-free-energy-around-saddle-point}
\eeq
with
\beq
\delta \Delta(\hat{z})=\Delta(\hat{z})-\Delta(\hat{z})^{*}
\eeq

For simplicity let us consider a glass state with spatially 
uniform density profile $\hat{\varphi}(\hat{z})=\hat{\varphi}$ and spatially uniform glass order parameter $\Delta(\hat{z})=\Delta$.
In this case the longitudinal Hessian becomes translationally invariant,
\beq
-M(\hat{z}_{1}-\hat{z}_{2})=-\frac{d}{2}(m-1)\hat{\varphi}
\left[
  \left(\frac{1}{\Delta^{2}}-\frac{m}{4}
  \hat{\varphi} X(\Delta)\right)\delta(\hat{z}_{1}-\hat{z}_{2})-\frac{m}{4}\hat{\varphi}X(\Delta)
  \frac{e^{-\frac{(\hat{z}_{1}-\hat{z}_{2})^{2}}{2}}}{\sqrt{2\pi}}
  \right].
\eeq
Introducing Fourier transforms as
\beq
M(\hat{z})=\int \frac{dk}{\sqrt{2\pi}}e^{ik\hat{z}}\hat{M}(k) \qquad
\delta \Delta(\hat{z})=\int \frac{dk}{\sqrt{2\pi}}e^{ik\hat{z}}\delta \Delta(k)
\eeq
we find the integral in the 2nd term in the r.h.s of \eq{eq-expansion-free-energy-around-saddle-point} becomes
\beq
\int d\hat{z}_{1}d\hat{z}_{2}M(\hat{z}_{1}-\hat{z}_{2})\delta \Delta(\hat{z}_{1})\delta \Delta(\hat{z}_{2})
=\int \frac{dk}{\sqrt{2\pi}}\tilde{M}(k)\delta \Delta(k)\delta \Delta(-k)
\eeq
with
\beq
\tilde{M}(k)=
\frac{d}{2}(m-1)\hat{\varphi} \left[
\left (
  \frac{1}{\Delta^{2}}-\frac{m}{4}\hat{\varphi}X(\Delta)
  \right)
  -
\frac{m}{4}\hat{\varphi}X(\Delta) 
  e^{-\frac{k^{2}}{2}} \right]
\eeq

Thus we find
\beq
\tilde{M}(k)=
\frac{d}{2}(m-1)\hat{\varphi}\left[M_{0}+\frac{k^{2}}{2}M_{2}+O(k^{4})\right]
\eeq
with
\beq
M_{0}=\frac{1}{\Delta^{2}}-\frac{m}{2}\hat{\varphi}X(\Delta) \qquad
M_{2}=\frac{m}{4}\hat{\varphi}X(\Delta)
\label{eq-m0-m2}
\eeq
Using these results we find the spatial correlation function of the fluctuation of
the glass order parameter as,
\beq
\langle \delta \Delta(\hat{z}_{1})\delta \Delta(\hat{z}_{2}) \rangle
\propto \exp \left(-\frac{|\hat{z}_{1}-\hat{z}_{2}|}{\xi^{\rm hessian}_{\rm d}}\right)
\eeq
with the correlation length $\xi_{\rm d}^{\rm hessian}$ given by
\beq
\xi^{\rm hessian}_{\rm d}=\sqrt{\frac{M_{2}}{2M_{0}}}
\label{eq-xi-dynamical-via-Hessian}
\eeq

\blue{Note that $M_{0}$ is nothing but the Hessian of the bulk system
(see \eq{eq-hessian-bulk}).}
Thus we must have
\beq
M_{0}=0 \qquad \qquad\mbox{at} \qquad\hat\varphi=\hat\varphi_{\rm d}
\eeq
We have also expected a scaling feature of $M_{0}$ approaching the dynamical transition point as \eq{eq-scaling-hessian-bulk} which reads
as
\blue{
\beq
M_{0} \propto \sqrt{\delta\hat\varphi}
\label{eq-scaling-M0}
\eeq
}
with $\delta\hat\varphi$ defined in \eq{eq-def-delta-hat-varphi} which measures the distance to the critical point.
This implies
\beq
\xi_{\rm d} \propto \delta\hat\varphi^{-1/4}.
\eeq

\begin{figure}[t]
\begin{center}
    \includegraphics[width=0.5\textwidth]{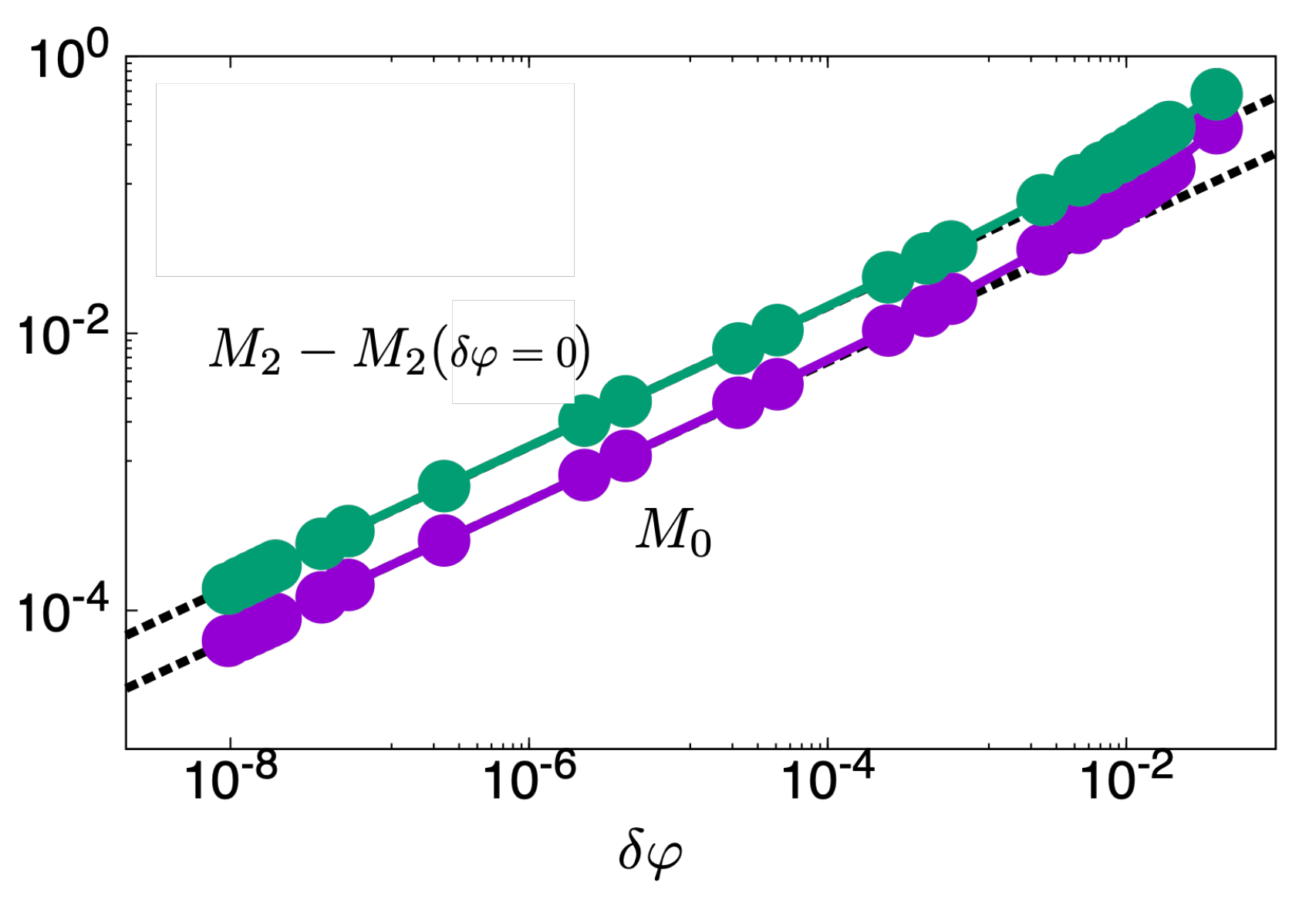}
\end{center}
\caption{
Scaling features of $M_{0}$ and $M_{2}$. Here the dotted lines are linear in $\delta\hat\varphi^{1/2}$.
}
\label{fig_m0_m2}
\end{figure}

Performing numerical analysis in the hard-sphere system for the case $m=1$, we indeed find
\beq
M_{0}=a\sqrt{\delta\hat\varphi} \qquad a\simeq 0.635
\qquad M_{2}=\frac{1}{2\Delta_{\rm d}^{2}} \simeq 0.376
\eeq
as shown in Fig.~\ref{fig_m0_m2}. 
Here $M_{2}=\frac{1}{2\Delta_{\rm d}^{2}}$ follows using $M_{0}=0$ which holds at the critical point
in \eq{eq-m0-m2} with $m=1$.

\subsection{Chemical potential and pressure}

In the present paper we will not investigate the ideal glass phase beyond the Kauzmann transition
where $0 < m < 1$.  As long as $m=1$, the chemical potential \eq{eq-chemical-potential-replica}
and the (reduced) pressure \eq{eq-reduced-pressure-replica}
become the same as those of the liquid given by 
\eq{eq-chemical-potential-liquid} and \eq{eq-reduced-pressure-liquid-theory}.
We will consider compression on the glassy metastable state
performing state following \cite{rainone2015following},
which amount to compress $m-1$ subset of replicas (see sec~\ref{sec-compression-replica})
in a subsequent work.

\subsection{Complexity}
\label{sec-complexity}

Let us analyze the complexity using the 1RSB free-energy \eq{eq-free-ene-1RSB-version1}
or \eq{eq-free-ene-1RSB}. To this end we have to examine first more closely the constant contribution to the entropic
part of the free-energy. We find
\beq
-\ln (\rho(\hat{z}))\lambda_{\rm th}^{d}=
  -\ln \left (\hat{\varphi}(\hat{z})(d/\Omega_{d})(\lambda_{\rm{th}}/D)^{d} \right) 
    =-\frac{1}{2}\ln{\qty(\pi \hat{\varphi}^{2}(\hat{z})d^{3})}
    +\frac{d}{2}\ln{\qty(\frac{2 \pi e D^2}{d^{2} \lambda_{\rm{th}}^{2}})}
    +\frac{d}{2}\ln{d}.
\eeq
Here we used
$\Omega_{d}=\pi^{\frac{d}{2}}/\Gamma(1+d/2)$ and $\Gamma(1+z)\sim\sqrt{2\pi z}\qty(z/e)^{z}$ ($z\gg 1$).
Using this we find
\beqn
   & \int d\hat{z} \hat\varphi(\hat{z})\qty{1-\ln{(\hat\varphi(\hat{z}) (d/\Omega_{d})(\lambda_{\rm{th}}/D)^{d})}
      + d\ln{m}+\frac{(m-1)d}{2}\ln{\frac{2\pi e D^2}{d^2\lambda_{\rm{th}}^{2}}}}  \nonumber \\
    &=\int d\hat{z} \hat\varphi(\hat{z})
    \qty{\qty(\frac{1}{2}-m)d\ln{d}
    +d\left\lbrack\frac{m}{2}\ln{\qty(\frac{2\pi e D^2}{\lambda_{\rm{th}}^2})}+\ln{m}\right\rbrack
    +\frac{1}{2}\ln{\qty(\frac{e^2}{\pi\hat{\varphi}^2(\hat{z}) d^3})}}.
\eeqn
Using the above expression we find the 1RSB free energy \eq{eq-free-ene-1RSB} as,
\begin{align}
    &\frac{-\beta F_{m} \qty[\qty{\Delta(\hat{z})}]}{S}\frac{\sqrt{d}}{D}\frac{\Omega_{d}D^{d}}{d}\nnum\\
    =&\int d\hat{z} \hat\varphi(\hat{z})
    \qty{\qty(\frac{1}{2}-m)d\ln{d}
    +d\left\lbrack\frac{m}{2}\ln{\qty(\frac{2\pi e D^2}{\lambda_{\rm{th}}^2})}+\ln{m}\right\rbrack
    +\frac{1}{2}\ln{\qty(\frac{e^2}{\pi\hat{\varphi}^{2}(\hat{z}) d^3})}
    +(-\beta m \hat{U}_{0}(\hat{z}))      
    }\nnum\\
    &
+\frac{d}{2}
\qty{
  \int d\hat{z} \hat\varphi({z})
  \qty[
2(-\beta m \hat{U}_{1}(\hat{z}))      
+
    (m-1)\ln{\frac{\Delta(\hat{z})}{2}}-\ln{m}]
    +
    \int d\hat{z}_{1} \hat\varphi(\hat{z}_{1})
    \int d\hat{z}_{2} \hat\varphi(\hat{z}_{2})
    \frac{e^{-\frac{(\hat{z}_{1}-\hat{z}_{2})^{2}}{2}}}{\sqrt{2\pi}}
    \qty(-\mathcal{F}_{\rm{int}}\qty(\Delta(\hat{z}_{1},\hat{z}_{2})))}
\end{align}

The complexity $\Sigma$ per molecule can be derived as $\Sigma^{*}=-m^{2}\partial_{m}\qty(-\frac{\beta F_{m}^{\rm 1RSB}}{mN})$
\cite{1130848328245474560}.
We have $N=(SD/\sqrt{d})\int d\hat{z} \rho(\hat{z})=(SD/\sqrt{d})\frac{d}{\Omega_{d}D^{d}}\int d\hat{z} \hat\varphi(\hat{z})$
(see \eq{eq-normalization-rho-mod}  and     \eq{eq-hat-var-phi}).
Thus we obtain the complexity per molecule as,
\beqn
    \label{append_complex_inhomogeneous}
    \Sigma^{*}
    &=&-m^{2}\partial_{m}\qty(-\frac{\beta F_{m}^{\rm{1RSB}}}{m} )\frac{\sqrt{d}}{SD}\frac{\Omega_{d}D^{d}}{d}
    \frac{1}{\int d\hat{z} \hat\varphi(\hat{z})}\nnum\\
    &=&\frac{1}{\int d\hat{z}\hat\varphi(\hat{z})} \left[\int d\hat{z} \hat\varphi(\hat{z})
    \qty{\frac{d}{2}\ln{d}
    -\frac{d}{2}\qty[1+\ln{\frac{\Delta(\hat{z})}{2m}}]}+O(d^{0})\right. \nnum\\
    &&+  \left. \frac{d}{2}
    \int d\hat{z}_{1} \hat\varphi(\hat{z}_{1})
    \int d\hat{z}_{2} \hat\varphi(\hat{z}_{2})
    \frac{e^{-\frac{(\hat{z}_{1}-\hat{z}_{2})^2}{2}} }{\sqrt{2\pi}}
    m^2\partial_{m}\qty(m^{-1}\mathcal{F}_{\rm{int}}\qty(\Delta_{ab}(\hat{z}_{1},\hat{z}_{2}))) \right]
\eeqn

\section{Cavity system}
\label{appendix-cavity}

Here we consider a cavity system of size $\hat{L}_{\rm{cav}}$ defined in the region $0 < \hat{z} < \hat{L}_{\rm{cav}}$.
The glass order parameter outside the cavity is set to zero
\beq
\Delta(\hat{z})=0\qquad -\infty < \hat{z}<0 \qquad \mbox{and}\qquad  \hat{L}_{\rm{cav}}<\hat{z} < \infty
\label{eq-cavity-condition}
\eeq
The density is set to be uniform $\hat\varphi(\hat{z})=\hat\varphi$ across the whole
system inside and outside the cavity.

In the cavity system the free-energy \eq{eq-F-large-d-glass-by-varphi}
becomes,
\beqn
&& \frac{-\beta F_{m}[\hat\varphi,q_{ab}]}{SD/\sqrt{d}}\frac{\Omega_{d}D^{d}}{d}=
\hat{L}_{\rm{cav}} \hat{\varphi}
\left [1-
  \ln \left (\hat{\varphi}(d/\Omega_{d})(\lambda_{\rm{th}}/D)^{d} \right)
 + d\ln{m}+\frac{(m-1)d}{2}\ln{\frac{2\pi e D^2}{d^2\lambda_{\rm{th}}^{2}}}
  \right]
\nonumber \\
&& +\frac{d}{2} \hat{\varphi}\left\{  \int_{0}^{\hat{L}_{\rm cav}} d\hat{z}
\ln \left(
    {\rm det}\;\hat{\alpha}^{m,m}(\hat{z})\right)
      \right. \nonumber \\
&& \left.    +\hat\varphi
    \qty[\int^{\infty}_{-\infty} d\hat{z}_{1}\int^{\infty}_{-\infty} d\hat{z}_{2}-\int_{\rm{ex-cav}} d\hat{z}_{1}\int_{\rm{ex-cav}} d\hat{z}_{2}]
\frac{e^{-\frac{(\hat{z}_{1}-\hat{z}_{2})^{2}}{2}}}{\sqrt{2\pi}}
\int d\xi e^{\xi}
e^{
-\frac{1}{2}  \sum_{a,b} 
     \Delta_{ab}(\hat{z}_{1},\hat{z}_{2})
       \frac{\partial^{2}}{\partial \xi_{a}\partial \xi_{b}}
}
 \left .
 f^{m}\left(
 \left \{
 D^{2}\left (1+ \frac{\xi_{a}}{d-1} \right)
\right \}
\right) \right |_{\xi^{a}=0}
\;\;\;
\right\}\nonumber \\
\label{eq-F-large-d-glass-by-varphi-cavity}
\eeqn
where
\begin{align}
    \int_{\rm{ex-cav}} d\hat{z}
    =\int^{\infty}_{-\infty} d\hat{z}
    -\int^{\hat{L}_{\rm{cav}}}_{0} d\hat{z}.
\end{align}
Note that the integrand in the double integral 
is symmetric with respect to the exchange of $\hat{z}_{1}$ and $\hat{z}_{2}$ so that we can replace the double integral by
\begin{align}
    \frac{1}{2}
    \qty[\int^{\infty}_{-\infty} d\hat{z}_{1}\int^{\infty}_{-\infty} d\hat{z}_{2}
    -\int_{\rm{ex-cav}} d\hat{z}_{1}\int_{\rm{ex-cav}} d\hat{z}_{2}]
    =\int_{-\infty}^{\infty}d\hat{z}_{1}\int_{0}^{\hat{L}_{\rm cav}}d\hat{z}_{2}
    -\frac{1}{2}  \int_{0}^{\hat{L}_{\rm cav}}d\hat{z}_{1}\int_{0}^{\hat{L}_{\rm cav}}d\hat{z}_{2}
    \label{eq-double-integral-conversion}
\end{align}
We see that in the last expression the first term represents the interaction between particles inside the cavity
and the second term represents that between particles inside and outside the cavity.
From the above expression we obtain the self-consistent equation for the glass order parameter in the cavity.

Assuming the 1RSB solution within the cavity, the saddle point equation
\eq{append_self_consistent_eq_most_general} becomes modified as,
\begin{align}
    \frac{1}{\Delta(\hat{z})}
    &=m\frac{\hat{\varphi} }{2}
    \blue{
    \qty[\int^{\infty}_{-\infty} d\hat{z}_{1}\int^{\infty}_{-\infty} d\hat{z}_{2}
    -\int_{\rm{ex-cav}} d\hat{z}_{1}\int_{\rm{ex-cav}} d\hat{z}_{2}]}    
    \frac{e^{-\frac{(\hat{z}_{1}-\hat{z}_{2})^2}{2}} }{\sqrt{2\pi}}
    \int_{-\infty}^{\infty} d\xi e^{\xi-\frac{1}{2}\Delta(\hat{z},\hat{z}')}g^{m}(\xi,\Delta(\hat{z},\hat{z}'))\qty(f'(\xi,\Delta(\hat{z},\hat{z}')))^{2}.
    \label{eq-SP-HS-1RSB-cavity}
\end{align}

Similarly the complexity \eq{append_complex_inhomogeneous} becomes,
\beqn
    \Sigma^{*}
    &=&-m^{2}\partial_{m}\qty(-\frac{\beta F_{m}^{\rm{1RSB}}}{m} )\frac{\sqrt{d}}{SD}\frac{\Omega_{d}D^{d}}{d}
    \frac{1}{\hat{L}_{\rm cav}\hat\varphi}\nnum\\
    &=&
\frac{1}{\hat{L}_{\rm cav}}
\left[\int_{0}^{\hat{L}_{\rm cav}} d\hat{z}
    \qty{\frac{d}{2}\ln{d}
    -\frac{d}{2}\qty[1+\ln{\frac{\Delta(\hat{z})}{2m}}]}+O(d^{0})\right. \nnum\\
    &&+  \left. \frac{d}{2}\hat\varphi
    \qty[\int^{\infty}_{-\infty} d\hat{z}_{1}\int^{\infty}_{-\infty} d\hat{z}_{2}
    -\int_{\rm{ex-cav}} d\hat{z}_{1}\int_{\rm{ex-cav}} d\hat{z}_{2}]    
    \frac{e^{-\frac{(\hat{z}_{1}-\hat{z}_{2})^2}{2}} }{\sqrt{2\pi}}
    m^2\partial_{m}\qty(m^{-1}\mathcal{F}_{\rm{int}}\qty(\Delta_{ab}(\hat{z}_{1},\hat{z}_{2}))) \right]
        \label{eq-complexity-cavity}
\eeqn

\section{Hard-sphere system}
\label{appendix-HS}

Here we collect some details for the hard-sphere system.

\subsection{Basics}

We consider hard-spheres whose interaction potential $v(r)$is given by
\beq
e^{-\beta v(r)}=\theta(r-D)
\eeq
where $D$ is the diameter with $\theta(x)$ being the Heaviside step function.
Using the scaled coordinate $\xi$ \eq{eq-xi} we find,
\beq
e^{-\beta v(D^{2}(1+\xi/d))}=\theta(\xi).
\eeq
Then the factor ${\cal F}$ defined by \eq{eq-cal-F} becomes
\begin{align}
     -\mathcal{F}
     =\int^{\infty}_{-\infty}d\xi e^{\xi}f\qty(D^{2}\qty(1+\frac{\xi}{d})^{2})
     =\int^{\infty}_{-\infty}d\xi e^{\xi}(\theta(\xi)-1)
     =1.
\end{align}

The function $g(\xi,\Delta)$ defined in \eq{eq-g-1RSB} becomes
\begin{align}
    g(\xi,\Delta)
    =\int \mathcal{D}w e^{-\beta v(D^2(1+\frac{\xi+\sqrt{\Delta}w}{d})^2)}
    =\int \mathcal{D}w\theta(\xi-\sqrt{\Delta} w)
    =\Theta\qty(\frac{\xi}{\sqrt{2\Delta}})
    \label{eq-g-HS}
\end{align}
where
\beq
\Theta(x) = \int_{-\infty}^{x}dz e^{-z^2}/\sqrt{\pi}=\qty(1+\erf(x))/2
    \label{eq-Theta}
\eeq
and $\erf(x)$ is the error function.
Similarly, $f'(\xi,\Delta)=-\partial_{\xi}\ln{g(\xi,\Delta)}$is,
\begin{align}
    -f'(\xi,\Delta)
    =\frac{\frac{e^{-\frac{\xi^2}{2\pi\Delta}}}{\sqrt{2\pi\Delta}}}{\Theta\qty(\frac{\xi}{\sqrt{2\Delta}})}
    =\frac{r\qty(\frac{\xi}{\sqrt{2\Delta}})}{\sqrt{2\Delta}},
    \label{eq-dashf-HS}
\end{align}
where we introduced
\beq
r(x)\equiv\Theta'(x)/\Theta(x)=e^{-x^2}/(\sqrt{\pi}\Theta(x))
\label{eq-r-HS}
\eeq
See Sec~\ref{append_subsection_assymptotic_behavior} for more details.

We obtain the self-consistent equation
of order parameter $\Delta(\hat{z})$ in hard-sphere system as,
\beqn
    \frac{1}{\Delta(\hat{z})}
    &=&\frac{m}{2}
    \int \frac{d\hat{z}'}{\sqrt{2\pi}}
    \hat{\varphi}(\hat{z}')
    e^{-\frac{(\hat{z}-\hat{z}')^2}{2}}
    \int_{-\infty}^{\infty} d\xi
    \frac{e^{\xi-\frac{1}{2}\Delta(\hat{z},\hat{z}')}}{2\Delta(\hat{z},\hat{z}')}
    \Theta^m\qty(\frac{\xi}{\sqrt{2\Delta(\hat{z},\hat{z}')}})
    r^2\qty(\frac{\xi}{\sqrt{2\Delta(\hat{z},\hat{z}')}}) \nonumber \\
   &=& \int \frac{d\hat{z}'}{\sqrt{2\pi}}
    \hat{\varphi}(\hat{z}')
    e^{-\frac{(\hat{z}-\hat{z}')^2}{2}}
    \int_{-\infty}^{\infty} d\xi
    \frac{1}{\Delta(\hat{z},\hat{z}')}\zeta_{m}(\Delta(\hat{z},\hat{z}')).
 \label{append_self_consistent_eq_most_general-HS}    
\eeqn
where we introduced,
\begin{align}
    \zeta_{m}(\Delta(\hat{z},\hat{z}'))
    \equiv 
    \frac{m}{4}\int d\xi e^{\xi-\frac{\Delta(\hat{z},\hat{z}')}{2}}
    \Theta^m\qty(\frac{\xi}{\sqrt{2\Delta(\hat{z},\hat{z}')}})r^2\qty(\frac{\xi}{\sqrt{2\Delta(\hat{z},\hat{z}')}}).
\end{align}

Let us consider here a very large density regime where $\Delta(\hat{z})$ will become very small. In such a regime
we find
\begin{align}
    \zeta_{m}(\Delta)
    \xrightarrow{\Delta\to0}I(m)\sqrt{\Delta},
    ~~~~~
    I(m)=\frac{m}{2\pi\sqrt{2}}\int_{-\infty}^{\infty}d y \Theta^{m-2}(y)e^{-2y^2}.
\end{align}
using this is in \eq{append_self_consistent_eq_most_general-HS} we find   
\begin{align}
    \frac{1}{\Delta(\hat{z})}
    &=\int \frac{d\hat{z}'}{\sqrt{2\pi}}
    \hat{\varphi}(\hat{z}')
    e^{-\frac{(\hat{z}-\hat{z}')^2}{2}}
    \frac{1}{\Delta(\hat{z},\hat{z}')}\zeta_{m}(\Delta(\hat{z},\hat{z}'))\nnum\\
    &\xrightarrow{\Delta(\hat{z})\to0}
    \int \frac{d\hat{z}'}{\sqrt{2\pi}}
    \hat{\varphi}(\hat{z}')
    e^{-\frac{(\hat{z}-\hat{z}')^2}{2}}
    \frac{1}{\Delta(\hat{z},\hat{z}')}I(m)\sqrt{\Delta(\hat{z},\hat{z}')}
    =\int \frac{d\hat{z}}{\sqrt{2\pi}}
    \hat{\varphi}(\hat{z}')e^{-\frac{(\hat{z}-\hat{z}')^2}{2}}
    \frac{I(m)}{\sqrt{\Delta(\hat{z},\hat{z}')}}.
\end{align}
This is consistent with 
\beq
\Delta\sim\hat{\varphi}^{-2}/I^{2}(m)
\eeq
which is known for the bulk system \cite{parisi2010mean}.

\subsection{Kauzmann transition in cavity}
\label{sec-kauzmann-transition-in-cavity}

Let us consider the complexity in the cavity system \eq{eq-complexity-cavity}
of the hard-sphere system at high densities where $\Delta(\hat{z}) \sim 0$.        
We find
\beqn
    \Sigma^{*}
    &=&-m^{2}\partial_{m}\qty(-\frac{\beta F_{m}^{\rm{1RSB}}}{m} )\frac{\sqrt{d}}{SD}\frac{\Omega_{d}D^{d}}{d}
    \frac{1}{\hat{L}_{\rm{cav}}}\nnum\\
    &=&\frac{1}{\hat{L}_{\rm{cav}}} \left[\int^{\hat{L}_{\rm{cav}}}_{0} 
    \qty{\frac{d}{2}\ln{d}
    -\frac{d}{2}\qty[1+\ln{\frac{\Delta(\hat{z})}{2m}}]}+O(d^{0})\right. \nnum\\
    &&+  \left. \frac{d}{2}\hat\varphi
    \qty[\int^{\infty}_{-\infty} d\hat{z}_{1}\int^{\infty}_{-\infty} d\hat{z}_{2}
    -\int_{\rm{ex-cav}} d\hat{z}_{1}\int_{\rm{ex-cav}} d\hat{z}_{2}]    
    \frac{e^{-\frac{(\hat{z}_{1}-\hat{z}_{2})^2}{2}} }{\sqrt{2\pi}}
    \sqrt{2\Delta(\hat{z}_{1},\hat{z}_{2})}
    \left[
      m^2\partial_{m}\qty(m^{-1} J_{\Delta(\hat{z},\hat{z}')}(m))
      \right] \right. \nonumber \\
    && - \left.  \left. \frac{d}{2}\hat\varphi
    \qty[\int^{\infty}_{-\infty} d\hat{z}_{1}\int^{\infty}_{-\infty} d\hat{z}_{2}
    -\int_{\rm{ex-cav}} d\hat{z}_{1}\int_{\rm{ex-cav}} d\hat{z}_{2}]    
    \frac{e^{-\frac{(\hat{z}_{1}-\hat{z}_{2})^2}{2}} }{\sqrt{2\pi}}
    \right.
    \right] \nonumber \\
    && \xrightarrow{\Delta \to 0}
    \frac{1}{\hat{L}_{\rm{cav}}} \left[\int^{\hat{L}_{\rm{cav}}}_{0} d\hat{z} 
    \qty{\frac{d}{2}\ln{d}
    -\frac{d}{2}\hat\varphi\qty[1+\ln{\frac{\Delta(\hat{z})}{2m}}]}+O(d^{0})\right. 
    \nonumber \\
    && - \left.  \left. \frac{d}{2}\hat\varphi
    \qty[\int^{\infty}_{-\infty} d\hat{z}_{1}\int^{\infty}_{-\infty} d\hat{z}_{2}
    -\int_{\rm{ex-cav}} d\hat{z}_{1}\int_{\rm{ex-cav}} d\hat{z}_{2}]    
    \frac{e^{-\frac{(\hat{z}_{1}-\hat{z}_{2})^2}{2}} }{\sqrt{2\pi}}
    \right.
    \right] \nonumber \\
  &&   \xrightarrow{d \gg 1}
    \frac{d}{2}\left( \ln{d}+\hat\varphi    \qty[2-f(\hat{L}_{\rm cav})  
    ]        \right)
\eeqn
where we introduced
\beq
J_\Delta(m)
=  \frac{1}{\sqrt{2\Delta}}\left[\mathcal{F}_{\rm{int}}\qty(\Delta)-1\right]
=  \frac{1}{\sqrt{2\Delta}}\int_{-\infty}^{\infty}d\xi e^{\xi} \left[ \Theta^{m}\left(\frac{\xi-\Delta/2}{\sqrt{2\Delta}}\right)-\theta(\xi) \right]\xrightarrow{\Delta \to 0}\int_{-\infty}^{\infty}dy[\Theta^{m}(y)-\theta(y)]
\eeq
and
\beq
f(\hat{L}_{\rm cav})=
\frac{1}{\hat{L}_{\rm cav}}\int_{0}^{\hat{L}_{\rm cav}} d\hat{z}_{1}
\int_{0}^{\hat{L}_{\rm cav}} d\hat{z}_{2}
\frac{e^{-\frac{(\hat{z}_{1}-\hat{z}_{2})^2}{2}} }{\sqrt{2\pi}}
=\erf\qty(\frac{\hat{L}_{\rm{cav}}}{\sqrt{2}})+\frac{1}{\hat{L}_{\rm{cav}}}\sqrt{\frac{2}{\pi}}\qty(e^{-\frac{\hat{L}_{\rm{cav}}^2}{2}}-1)
=1-\sqrt{\frac{2}{\pi}}\hat{L}_{\rm{cav}}^{-1}+O\qty(e^{-\hat{L}_{\rm{cav}}^{2}/2}).
\eeq
To derive the above results we also used \eq{eq-double-integral-conversion}, \eq{eq-f-cav-formula} and  \eq{eq-function-f-assymptotic}.

Therefore, the Kauzmann transition density $\hat{\varphi}_{K}(\hat{L}_{\rm{cav}})$
of the cavity system at which the complexity vanishes is obtained as,
\begin{align}
    \hat{\varphi}_{K}(\hat{L}_{\rm{cav}})
    =\frac{\ln{d}}{2-f(\hat{L}_{\rm{cav}})}
    =\frac{\hat{\varphi}_{K,\rm{bulk}}}{2-f(\hat{L}_{\rm{cav}})} \simeq 
    \frac{\hat{\varphi}_{K,\rm{bulk}}}{1-\sqrt{\frac{2}{\pi}}\hat{L}_{\rm{cav}}^{-1}}.
\end{align}
where $\hat{\varphi}_{K,\rm{bulk}}$ is the Kauzmann transition density for the bulk system $\hat{L}_{\rm cav}=\infty$.
Thus in cavity systems the Kauzmann transition occurs at lower densities than in bulk systems and
the transition density increases increasing the cavity size $\hat{L}_{\rm cav}$.

We obtain the PS length at the Kauzmann transition,
\begin{align}
    \xi_{K}
    =\frac{\hat{L}_{\rm{cav}}}{2}
    =\sqrt{\frac{2}{\pi}}\frac{\hat{\varphi}_{K}(\hat{L}_{\rm{cav}})}{\hat{\varphi}_{K,\rm{bulk}}-\hat{\varphi}_{K}(\hat{L}_{\rm{cav}})}
    \propto(\hat{\varphi}_{K,\rm{bulk}}-\hat{\varphi}_{K}(\hat{L}_{\rm{cav}}))^{-1}.
\end{align}
Thus, the exponent of the PS length at the Kauzmann transition is $-1$.

\section{Useful Formulas}
    \label{append_useful_equations}

The following formula can be proved by taking the direct derivative.
\begin{align}
    \left.\sum_{a=1}^{n}\pdv{}{h_{a}}\prod_{c=1}^{n}f(h_{c})\right|_{\qty{h_{a}=h}}
    &=\pdv{}{h}f^{n}(h)\nnum\\
    \left.\sum_{a=1}^{n}\pdv{}{h_{a}}{h_{b}}\prod_{c=1}^{n}f(h_{c})\right|_{\qty{h_{a}=h}}
    &=\pdv[2]{}{h}f^{n}(h)
    \label{eq-diff-formulae}
\end{align}

The following formula can be proved with the Taylor expansion $f(h+\delta)=\sum_{n=0}^{\infty}\frac{\delta^{n}}{n!}\partial_{h}^{n}f(h)$.
\begin{align}
  e^{\frac{a}{2}\pdv[2]{}{h}}f(h)
= \int \mathcal{D}ze^{-\frac{z^{2}}{2}}f(h+\sqrt{a}z)
    \label{eq-diff-gaussian-integral-conversion}
\end{align}
where we introduced a short-hand notation,
\beq
\int \mathcal{D} z = \int^{\infty}_{-\infty}\frac{dz}{\sqrt{2\pi}}\blue{e^{-\frac{z^{2}}{2}}}.
\label{eq-def-Dz}
\eeq

\subsection{Asymptotic behavior of the error function}
    \label{append_subsection_assymptotic_behavior}

The err function $\erf(x)$ is an odd function,
\begin{align}
    \erf(x)
    =\frac{2}{\sqrt{\pi}}
    \int_{0}^{x}dt e^{-t^2}
    =-\erf(-x).
\end{align}

The behavior of the error function at $x\to\infty$ is
\begin{align}
    \label{append_erf_assymptotic_behavior}
    \erf(x)
    =1-\frac{1}{\sqrt{\pi}}\frac{e^{-x^{2}}}{x}
    \qty(1-\frac{1}{2x^{2}}+\frac{3}{(2x^{2})^{2}}+\cdots).
\end{align}

This can be proved as follows.
\begin{align}
    \erf(x)
    =\frac{2}{\sqrt{\pi}}\int_{0}^{x}dt e^{-t^2}
    =\frac{2}{\sqrt{\pi}}\qty(\int_{0}^{\infty}dt e^{-t^2}-\int_{x}^{\infty}dt e^{-t^2})
    =1-\frac{2}{\sqrt{\pi}}
    \int_{x}^{\infty}dt e^{-t^2}
\end{align}
With $\partial_{x}e^{-x^2}=-2x e^{-x^2}$, for the second term,
\begin{align}
    &\int_{x}^{\infty}dt e^{-t^2}
    =\int_{x}^{\infty}dt \frac{-t^{-1}}{2}\partial_{t}e^{-t^2}\nnum\\
    =&\qty[\frac{-t^{-1}e^{-t^2}}{2}]_{x}^{\infty}-\int_{x}^{\infty}dt \frac{t^{-2}e^{-t^2}}{2}
    =\frac{x^{-1}e^{-x^2}}{2}
    +\int_{x}^{\infty}dt \frac{t^{-3}}{4}\partial_{t}e^{-t^2}\nnum\\
    =&\frac{x^{-1}e^{-x^2}}{2}
    -\frac{x^{-3}e^{-x^2}}{4}
    +\int_{x}^{\infty}dt\frac{3t^{-4}}{4}e^{-t^2}
    =\frac{x^{-1}e^{-x^2}}{2}
    -\frac{x^{-3}e^{-x^2}}{4}
    -\int_{x}^{\infty}dt\frac{3t^{-5}}{8}\partial_{t}e^{-t^2}\nnum\\
    =&e^{-x^2}\qty[\frac{x^{-1}}{2}-\frac{x^{-3}}{4}+\frac{3x^{-5}}{8}+O\qty(x^{-7})]
    =\frac{e^{-x^2}}{2x}\qty[1-\frac{1}{2x^{2}}+\frac{3}{(2x^{2})^{2}}+O\qty(x^{-7})].
\end{align}

Using the above equation, the behavior of $\Theta(x)$ at $x\to\infty$ is,
\begin{align}
    \Theta(x)
    =\int^{x}_{-\infty}\frac{dz}{\sqrt{\pi}}e^{-z^{2}}
    =\frac{1+\erf(x)}{2}
    =\left\{
    \begin{array}{ll}
        \frac{1}{2}\frac{e^{-x^{2}}}{(-x)\sqrt{\pi}}\qty[1-\frac{1}{2x^{2}}+\frac{3}{(2x^{2})^{2}}+\cdots] & x\to-\infty \\
        ~~~~~~~~~~~~~~~~~~~1 & x\to\infty
    \end{array}
    \right. .
\end{align}

And,
\begin{align}
    r(x)
    =\frac{\Theta'(x)}{\Theta(x)}
    =\frac{e^{-x^{2}}}{\sqrt{\pi}\Theta(x)}
\end{align}
behaves asymptotically like 
\begin{align}
    r(x)
    =\left\{
    \begin{array}{ll}
        -2x\qty[1-\frac{1}{2x^{2}}+\frac{3}{(2x^{2})^{2}}+\cdots]^{-1} & x\to-\infty \\
        ~~~~~~~~~~~~~0 & x\to\infty
    \end{array}
    \label{eq-r-HS-asymptotic}
    \right. .
\end{align}
When performing numerical calculations using $\Theta(x)$ or $r(x)$, you can use these asymptotic expressions.

\subsection{The function of $f(L)$}
\label{append_subsection_function_of_complexity}

We derive
\begin{align}
    f(\hat{L}_{\rm{cav}})
    =\frac{1}{\hat{L}_{\rm{cav}}}\int^{\hat{L}_{\rm{cav}}}_{0}d\hat{z}\int^{\hat{L}_{\rm{cav}}}_{0} d\hat{z}'
    \frac{1}{\sqrt{2\pi}}
    e^{-\frac{(\hat{z}-\hat{z}')^2}{2}}
    =\erf\qty(\frac{\hat{L}_{\rm{cav}}}{\sqrt{2}})+\frac{1}{\hat{L}_{\rm{cav}}}\sqrt{\frac{2}{\pi}}\qty(e^{-\frac{\hat{L}_{\rm{cav}}^2}{2}}-1).
    \label{eq-f-cav-formula}
\end{align}

For the err function $\erf(x)$,
\begin{align}
    \int^{L}_{0}dz
    \erf\qty(\frac{z}{\sqrt{2}})
    &=\int^{L}_{0}dz \qty(\partial_{z}z)
    \erf\qty(\frac{z}{\sqrt{2}})\nnum\\
    &=\int^{L}_{0}dz \partial_{z}\qty[z\erf\qty(\frac{z}{\sqrt{2}})]
    -\sqrt{\frac{2}{\pi}}
    \int^{L}_{0}dz z e^{-\frac{z^2}{2}}\nnum\\
    &=L\erf\qty(\frac{L}{\sqrt{2}})
    +\sqrt{\frac{2}{\pi}}\qty(e^{-\frac{L^2}{2}}-1).
\end{align}
With this,we can derive
\begin{align}
    \int^{L}_{0}dz
    \int^{L}_{0}dw
    e^{-\frac{(z-w)^2}{2}}
    &=\int^{L}_{0}dz\int^{L}_{0}dw
    \partial_{w}
    \qty[\sqrt{\frac{\pi}{2}}\erf\qty(\frac{w-z}{\sqrt{2}})]\nnum\\
    &=\sqrt{\frac{\pi}{2}}\int^{L}_{0}dz
    \qty[\erf\qty(\frac{L-z}{\sqrt{2}})
    +\erf\qty(\frac{z}{\sqrt{2}})]\nnum\\
    &=\sqrt{2\pi}\int^{L}_{0}dz
    \erf\qty(\frac{z}{\sqrt{2}})\nnum\\
    &=\sqrt{2\pi}L\qty[\erf\qty(\frac{L}{\sqrt{2}})+\frac{1}{L}\sqrt{\frac{2}{\pi}}\qty(e^{-\frac{L^2}{2}}-1)]
\end{align}

In addition, the asymptotic behavior of
\begin{align}
    f(x)
    =\erf\qty(\frac{x}{\sqrt{2}})
    +\frac{1}{x}\sqrt{\frac{2}{\pi}}\qty(e^{-\frac{x^2}{2}}-1)
\end{align}
in $x\to\infty$ is
\begin{align}
    f(x)
    &=1-\frac{1}{x}\sqrt{\frac{2}{\pi}}e^{-\frac{x^2}{2}}\qty(1-x^{-2}+3x^{-4}+O(x^{-6}))+\frac{1}{x}\sqrt{\frac{2}{\pi}}\qty(e^{-\frac{x^2}{2}}-1)\nnum\\
    &=1-\sqrt{\frac{2}{\pi}}x^{-1}+O\qty(x^{-3}e^{-\frac{x^2}{2}})
    \label{eq-function-f-assymptotic}
\end{align}
from Eq.\ref{append_erf_assymptotic_behavior}.

\section{Collection of some useful formulas}

\subsection{Proof of \texorpdfstring{$\Omega_{d-1}=\sqrt{d/2\pi}\Omega_{d}$}{TEXT}}
    \label{append_omega_proof}

The $\Omega_{d}$ is the volume of a hyper-sphere of radius $1$ in $d$-dimension,
\begin{align}
    \Omega_{d}=\frac{\pi^{\frac{d}{2}}}{\Gamma\qty(\frac{d}{2}+1)}.
\end{align}
Here, $\Gamma$ if gamma function. 
With the Stirling's approximation, 
\begin{align}
    \Gamma(n+1)
    =n! \sim \sqrt{2\pi n}\qty(\frac{n}{e})^n
\end{align}
For sufficiently large $d$,
\begin{align}
    \frac{\Omega_{d}}{\Omega_{d-1}}
    =\sqrt{\pi}\frac{\Gamma(\frac{d-1}{2}+1)}{\Gamma(\frac{d}{2}+1)}
    =\sqrt{\frac{2\pi}{d}}.
\end{align}
because
\begin{align}
    \frac{\Gamma(\frac{d-1}{2}+1)}{\Gamma(\frac{d}{2}+1)}
    =\frac{\sqrt{\frac{d-1}{2}}\qty(\frac{d-1}{2e})^{\frac{d-1}{2}}}{\sqrt{\frac{d}{2}}\qty(\frac{d}{2e})^{\frac{d}{2}}}
    =\qty(1-\frac{1}{d})^{\frac{1}{2}}
    \qty(\frac{2e}{d-1})^{\frac{1}{2}}
    \qty(1-\frac{1}{d})^{\frac{d}{2}}
    =\sqrt{\frac{2e}{d}}\qty(1-\frac{1}{d})^{\frac{d}{2}}=\sqrt{\frac{2}{d}}.
\end{align}
Thus we get 
\begin{align}
    \Omega_{d-1}=\sqrt{\frac{d}{2\pi}}\Omega_{d}.
\end{align}

\subsection{Integration of Gaussians}
    \label{append_gaussian_proof}

The integral of a Gaussian is as follows.

\begin{align}
    \int\qty[\prod_{i=1}^{M}dx_{i}]
    e^{-\frac{1}{2}\sum_{i,j}x_{i}K_{ij}x_{j}}
    =\frac{{\sqrt{2\pi}}^{M}}{\sqrt{\det K}}
\end{align}

\begin{align}
    \frac{\sqrt{\det K}}{{\sqrt{2\pi}}^{M}}
    \int\qty[\prod_{i=1}^{M}dx_{i}]e^{-\frac{1}{2}\sum_{i,j}x_{i}K_{ij}x_{j}+\sum_{i}h_{i}x_{i}}
    =e^{\frac{1}{2}\sum_{i,j}h_{i}K_{ij}^{-1}h_{j}}
\end{align}

\section{\blue{Numerical details to solve the integral equation \eq{eq-SP-HS-1RSB-cavity}}}
\label{sec-numerical-detail-how-to-solve-SP}

\blue{
We solved the integral equation  \eq{eq-SP-HS-1RSB-cavity} to obtain the space dependent order
parameter $\Delta(\hat{z})$ for the system of hard-spheres in the cavity as follows.
\begin{enumerate}
\item The cavity region $0 < \hat{z} < \hat{L}_{\rm cav}$ is divided into segments of width 
$d \hat{z}=0.1$ for each of which we evaluate $\Delta(\hat{z})$ as follows.
\item The integrations over $\hat{z}'$ and $\xi$ are done using the simple Euler method
with $d \hat{z}'=d\xi=0.1$. For the evaluation of the function $g(\xi,\Delta)$ we used the expression
\eq{eq-g-HS} with \eq{eq-Theta}. For the evaluation of the function $f'(\xi,\Delta)$,
we used \eq{eq-dashf-HS}, \eq{eq-r-HS} and \eq{eq-r-HS-asymptotic}.
\item Let us call the initial guess for $\Delta(\hat{z})$ (see 6)
  as $\Delta(\hat{z};t=0)$.
  \item We evaluate the r.h.s of
  \eq{eq-SP-HS-1RSB-cavity} using $\Delta(\hat{z};t)$ to obtain $1/\Delta(\hat{z};t+1)$.
\item The procedure 4 is repeated for the iteration steps $t=0,1,2,\ldots$ until
  the order parameters converge.
\item  For the initial value of $\Delta(\hat{z},t=0)$, we used
  the solution of the self-consistent equation
  \eq{append_self_consistent_eq_most_uniform} for the bulk system with $m=1$.
  \eq{append_self_consistent_eq_most_uniform} is also solved iteratively until the solution converges.
\end{enumerate}
}

\newpage
\bibliographystyle{unsrt}
\bibliography{tomita_yoshino.v2}
\end{document}